\documentclass[multphys,vecphys]{svmult}

\usepackage{makeidx}     
\usepackage{graphicx}    
\usepackage{multicol}    
\usepackage{amssymb}     

\makeindex             

\newcommand{\half}{\frac{1}{2}}

\newcommand{\etal}{{\it et al.}}

\newcommand{\ie}{{\em i.e.~~}}
\newcommand {\lleq} {\lower0.9ex\hbox{ $\buildrel < \over \sim$} ~}
\newcommand{\ggeq} {\lower0.9ex\hbox{ $\buildrel > \over \sim$} ~}
\newcommand{\tik} {\langle T_{ik}\rangle}
\newcommand\statei{\lbrace r,s \rbrace}
\newcommand\atridot{\stackrel{...}{a}}
\newcommand{\omx} {\Omega_X}
\newcommand{\n}{\noindent}
\newcommand{\om}{\Omega}
\newcommand{\la}{\Lambda}

\newcommand{\apj}{{\it Astroph.~J.~}}
\newcommand{\mn}{{\it Mon.~Not. Roy.~ast. Soc.~}}
\newcommand{\asta}{{\it Astron.~Astrophys.~}}

\newcommand{\prl}{{\it Phys.~Rev. Lett.~}}
\newcommand{\pd}{{\it Phys.~Rev.~D~}}

\newcommand{\plb} {{\it Phys.~Lett.~B~}}
\newcommand{\jetpl} {JETP ~Lett.~}

\newcommand{\beq}{\begin{equation}}
\newcommand{\eeq}{\end{equation}}
\newcommand{\ber}{\begin{eqnarray}}
\newcommand{\eer}{\end{eqnarray}}

\newcommand{\sq}{\lower.25ex\hbox{\large$\Box$}}

\begin{document}

\title*{Dark Matter and Dark Energy}
\author{
Varun Sahni}

\institute{Inter-University Centre for Astronomy \& Astrophysics,\\
Pun\'e 411 007, India\\
\texttt{varun@iucaa.ernet.in}}

\maketitle

\begin{abstract}
I briefly review our current understanding of dark matter and dark energy.
The first part of this paper focusses on issues pertaining to dark matter
including observational evidence for its existence, current constraints
and the `abundance of substructure' and `cuspy core' issues which arise in CDM.
I also briefly describe MOND.
The second part of this review focusses on dark energy.
In this part I discuss the significance of the cosmological constant problem
which leads to a predicted value of the cosmological constant which is 
almost $10^{123}$ times larger than the observed value 
$\la/8\pi G \simeq 10^{-47}$GeV$^4$. 
Setting $\la$ to this small value ensures that
the acceleration of the universe is a 
fairly recent phenomenon giving rise to 
the `cosmic coincidence' conundrum 
according to which we live 
during a special epoch when the density in matter and $\la$ are almost equal.
Anthropic arguments are briefly discussed but more emphasis is placed upon
dynamical dark energy models in which the equation of state is time dependent.
These include Quintessence, Braneworld models,
Chaplygin gas and Phantom energy. 
Model independent methods to determine the cosmic equation of state and
the Statefinder diagnostic are also discussed. 
The Statefinder has the attractive property
$\atridot/a H^3 = 1 $ for LCDM, which is helpful for differentiating
between LCDM and rival dark energy models.
The review ends with a brief discussion of the fate of
the universe in dark energy models.
\end{abstract}

\newpage

\section{Dark Matter}
\index{dark matter}

Observations of the cosmic microwave background (CMB) and 
the deuterium abundance in the Universe suggest that
$\om_{\rm baryon}h^2 \simeq 0.02$, or $\om_{\rm baryon} \simeq 0.04$ 
if the current Hubble expansion rate is $h = H_0/100
{\rm km/sec/Mpc} = 0.7$.
Although $\om_{\rm baryon}$ is
much larger than the observed mass in stars, $\om_{\rm stars} \simeq 0.005$
\footnote{This suggests that most of the baryonic matter at 
$z=0$ is not contained in stars
but might be contained in hot gas \cite{bosma03}.}, it
is nevertheless very much smaller than the total energy density in the universe 
inferred from the
observed anisotropy in the cosmic microwave background \cite{spergel03}
\beq
\Omega_{\rm total} \equiv \frac{8\pi G\rho_{\rm total}}{3H^2} = 
1.02 \pm 0.02~.
\eeq
Both dark matter and dark energy are considered essential missing pieces
in the cosmic jigsaw puzzle
\beq
\om_{\rm total} - \om_{\rm baryons} = ~?
\eeq
Although the nature of neither dark matter (DM) nor dark energy (DE)
is currently known, it is felt that both DM and DE are non-baryonic in origin,
and that DM is distinguished from
DE by the fact that the former clusters on sub-Megaparsec scales
(in order to explain galactic rotation curves) whereas the latter
has a large negative pressure (and can make the universe accelerate).
In addition there is strong evidence to suggest that 
\beq
\Omega_m \simeq 1/3, ~~ \Omega_{\rm DE} \simeq 2/3~.
\eeq

\n
In these lectures I will briefly review some properties of both
dark matter and dark energy.

Though the observational evidence favouring a flat Universe with $\Omega_{\rm total} 
\simeq 1$ is fairly 
recent, the nature of the `unseen' component of the universe (which 
dominates its mass density), is a long-standing issue in
modern cosmology.  
Indeed, the need for dark matter was originally pointed out by Zwicky (1933)
who realized that the velocities of individual galaxies located within the Coma cluster
were quite large, and that this cluster would be gravitationally bound only
if its total mass substantially exceeded the sum of the masses of its component galaxies.
For clusters which have relaxed to dynamical equilibrium 
the mean kinetic and potential energies are related by the 
virial theorem \cite{coles}
\beq
K + \frac{U}{2} = 0~,
\eeq
where $U \simeq -GM^2/R$ is the potential energy of a cluster of radius $R$, 
$K \simeq 3 M\langle v_r^2\rangle/2$ is the kinetic energy and
$\langle v_r^2\rangle^{1/2}$ is the dispersion in the line-of-sight velocity 
of cluster galaxies. 
(Clusters in the Abell catalogue typically have $R \simeq 1.5 h^{-1}$ Mpc.)
This relation allows us to infer the mean gravitational potential energy if 
the kinetic energy is accurately known.
The mass-to-light ratio in clusters
can be as large as $M/L \simeq 300 M_\odot/L_\odot$. 
However since most of the mass in clusters is in the form of hot, x-ray emitting intracluster
gas, the extent of dark matter in these objects is estimated to be 
$M/M_{\rm lum} \simeq 20$, where $M_{\rm lum}$ is the total mass in luminous matter
including stars and gas. 

\begin{figure}[ht]
\centering
\includegraphics[width=10cm]{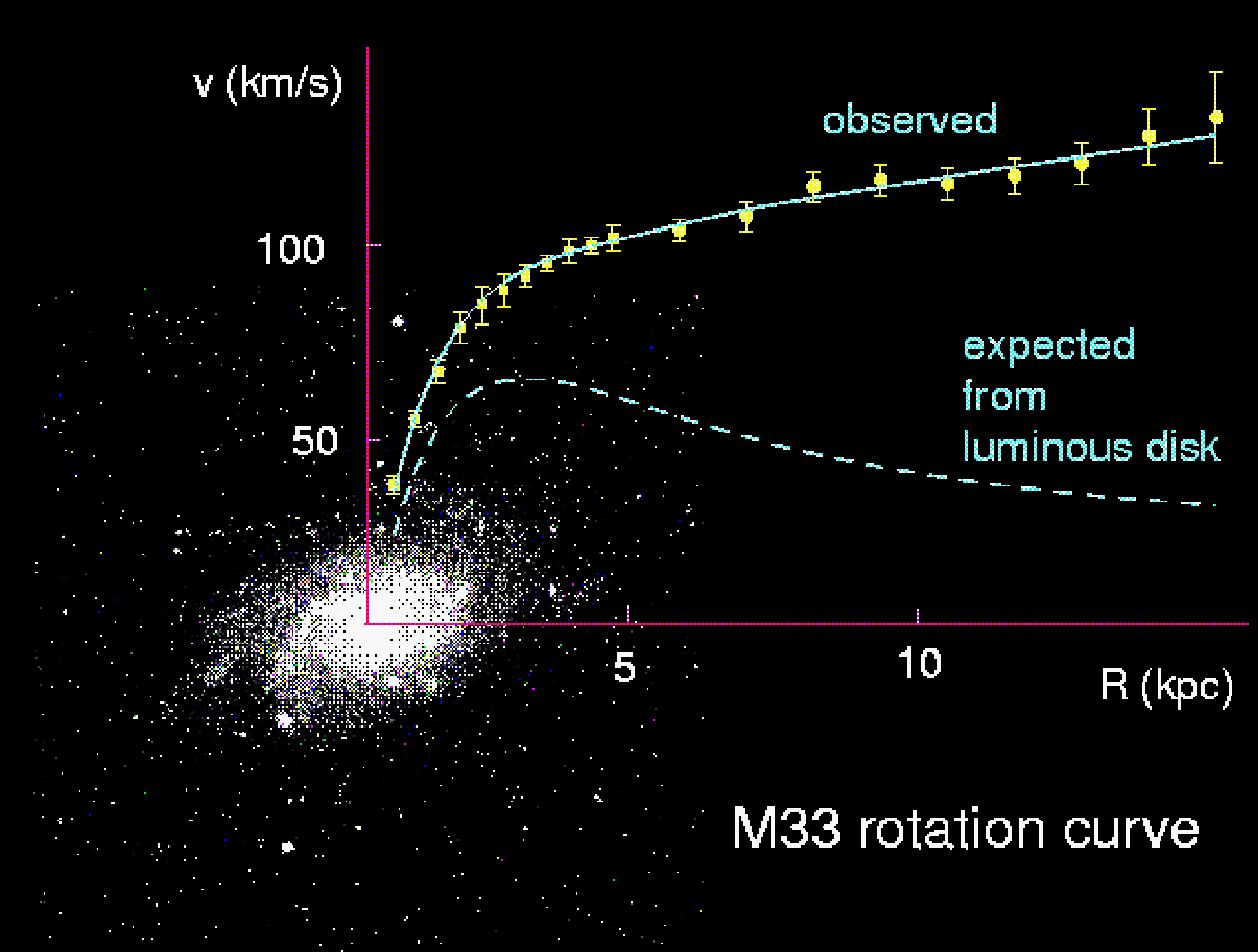}
\caption{\footnotesize
The observed rotation curve of the dwarf spiral galaxy M33 extends considerably
beyond its optical image (shown superimposed); 
from Roy ~\protect\cite{dproy}.
}
\label{fig:flat}
\end{figure}

In individual galaxies the presence of dark matter has been convincingly established
through the use of Kepler's third law
\beq
v(r) = \sqrt{\frac{G M(r)}{r}}
\eeq
to determine the `rotation curve' $v(r)$ at a given radial distance from the galactic
center.  Observations of galaxies taken at distances large enough
for there to be no luminous galactic component indicate that, instead of 
declining at the expected rate $v \propto r^{-1/2}$ true if $M \simeq {\rm constant}$, the velocity curves 
flattened out to $v \simeq {\rm constant}$ implying $M(r) \propto r$ (see fig \ref{fig:flat}).
This observation suggests that the mass of galaxies continues to grow even
when there is no luminous component to account for this increase.
Velocity curves have been compiled for over 1000 spiral galaxies usually by
measuring the 21 cm emission line from neutral hydrogen (HI) \cite{dark1,dark2}.
The results indicate that $M/L = (10 - 20) M_\odot/L_\odot$ in spiral
galaxies and in ellipticals, while this ratio can increase
to $M/L \simeq (200 - 600) M_\odot/L_\odot$ in low surface brightness galaxies
(LSB's) and in dwarfs. 
For instance, a recent measurement of the Draco dwarf spheroidal galaxy
located 
at a distance of only 79 kpc from the Milky Way shows the presence of a considerable
amount of dark matter 
 $M/L\vert_{\rm Draco} = 440 \pm 240 M_\odot/L_\odot$ \cite{kleina01} !
It is interesting that the total mass of an individual galaxy
is still somewhat of an unknown quantity since
a turn around to the $v \propto r^{-1/2}$ law at large radii has not been convincingly
observed.

An important difference between the distribution of dark matter in galaxies and
clusters needs to be emphasised: whereas dark matter
appears to {\em increase} with distance in galaxies, in clusters exactly the reverse
is true, the dark matter distribution actually {\em decreases} with distance.
Indeed, for certain dwarfs (such as DD0154) the rotation curve has been measured to
almost 15 optical length scales indicating that the dark matter surrounding this object
is extremely spread out (see also figure \ref{fig:flat}). 
A foreground cluster, on the other hand,
acts as a gravitational lens which focuses the light from
background objects such as galaxies and QSO's thereby allowing us to determine
the depth of the cluster potential well. Observations of strong lensing by clusters
indicate that dark matter is strongly concentrated in central regions with a projected
mass of $10^{13} - 10^{14} M_\odot$ being contained within 0.2 - 0.3 Mpc of the central region.
As we shall see later, this observation may prove to be problematic for alternatives
to the dark matter hypothesis such as the Modified Newtonian Dynamics (MOND) approach of
Milgrom \cite{milgrom}.

As discussed earlier, the fact that only $4\%$ of the cosmic density is
baryonic suggests that the dark matter which we are observing could well be
non-baryonic in origin.
The need for non-baryonic forms of dark matter gets indirect  support from the fact
that baryonic models find it difficult to grow structure from small initial
conditions and hence to reconcile
the existence of a well developed cosmic web of filaments, sheets and clusters
at the present epoch with the exceedingly small amplitude of density perturbations
($\delta\rho/\rho \sim 10^{-5}$ at $z \simeq 1,100$)
inferred from COBE measurements and more recent
CMB experiments.
Indeed, it is well known that, if the effects of pressure are ignored,
linearized density perturbations in a spatially flat matter dominated universe 
grow at the rate $\delta \propto t^{2/3} \propto (1+z)^{-1}$, 
where $1+z = a_0/a(t)$ is the cosmological
redshift. 
(Contrast this relatively slow growth rate with the exponential `Jeans instability' of a static
matter distribution $\delta \propto \exp{\sqrt{4\pi G\rho} t}$.)
In a baryonic universe, the large radiation pressure (caused by thompson scattering of
CMB photons off electrons) ensures that density perturbations in the
baryonic component can begin growing only after
hydrogen recombines at $z \simeq 1,100$ at which point of time 
baryons and radiation decouple.
Requiring $\delta > 1$ today implies $\delta > 10^{-3}$ at
recombination, which contradicts CMB observations by over an order of magnitude !
In non-baryonic models on the other hand, 
the absence of any significant coupling between dark matter
and radiation allows structure to grow
much earlier, significantly before hydrogen in the universe has recombined. After recombination
baryons simply fall into the potential wells created for them by the dominant 
non-baryonic component. As a result
a universe with a substantial non-baryonic component
can give rise to the structure which we see today from
smaller initial fluctuations.

The growth of structure via gravitational instability depends both upon the
nature of primordial perturbations (adiabatic/isocurvature) and upon whether
the dark matter species is {\rm hot} or {\rm cold}.
The issue of density perturbations has been discussed in considerable detail
by Ruth Durrer at this school and I shall not touch upon this important topic
any further. Let me instead say a few words about 
hot and cold dark matter.
Non-baryonic {\em Hot Dark Matter} (HDM) particles are assumed to have
decoupled from the rest of matter/radiation when they were relativistic
and so have a very
large velocity dispersion (hence called `hot'). 
{\em Cold Dark Matter} (CDM) particles,
on the other hand, have a very
small velocity dispersion and decouple from the rest of matter/radiation when
they are non-relativistic.
The free-streaming (collisionless phase mixing) of non-baryonic
 particles as they travel from high density to low density
regions (and vice versa) introduces an important length scale called 
the `free-streaming distance'
$\lambda_{\rm fs}$ -- which is the mean distance travelled by a relativistic
particle species until its momentum becomes non-relativistic.
In both HDM and CDM the processed {\em final} spectrum of density perturbations
differs from its initial form. In the case of HDM this difference arises because
fluctuations on scales smaller than $\lambda_{\rm fs}$ are wiped out due to free
streaming with the result that the processed final spectrum has a well defined
cutoff on scales smaller than $\lambda \sim \lambda_{\rm fs}$.
Perhaps the best example of HDM is provided by a light neutrino of mass about
$30$ eV. In this case $\lambda_{\rm fs} \simeq 41(30 {\rm eV}/m_\nu)$ Mpc
$\Rightarrow$ large proto-pancakes having masses comparable to those of
rich clusters of galaxies $M \sim 10^{15} M_\odot$ are the first objects to form
in HDM. Smaller objects (galaxies) are formed by the fragmentation of the 
proto-pancake. This {\em top-down} scenario for structure formation was originally
suggested by Zeldovich and coworkers in connection with adiabatic baryonic models
and subsequently applied to HDM. It has since fallen out of favour
mainly due to the strong observational constraints on the mass of the neutrino
$\sum_{\nu_i}m_{\nu_i} < 0.7$ eV 
and on the relic neutrino density $10^{-3} \lleq \Omega_\nu h^2 \lleq 10^{-1}$ 
\cite{elgaroy02,spergel03,ellis03,minaka02}.
It also faces considerable difficulty in forming structure 
sufficiently early to explain the existence of
galaxies and QSO's at high redshifts.

In contrast to HDM, constituents of
CDM have a much smaller
free-streaming distance. 
Because of this small scales are the first to go non-linear
and gravitational clustering proceeds in a {\em bottom up} fashion in this scenario.
A key quantity defining gravitational clustering
is the power spectrum of density perturbations
$P(k) \equiv |\delta_k|^2$, which is related to the mean square density fluctuation via
\beq
\bigg\langle \left (\frac{\delta\rho}{\rho}\right )^2 \bigg\rangle
= 4\pi \int_0^\infty P(k) k^2 dk~.
\label{eq:pow1}
\eeq
Inflationary models predict $P_i(k) \propto k^n$, $n \simeq 1$, at an early epoch.
As the universe expands the power spectrum gets modified.
The `processed' final spectrum depends upon the nature of dark matter, the epoch of 
matter-radiation equality and other cosmological quantities. The final and initial
spectra are related through a transfer function 
\beq
P_f(k) = P_i(k)\times T^2(k)~.
\label{eq:pow_spec}
\eeq
CDM-type spectra have the following approximate form of
the transfer function \cite{sahni84,ss84,sc95}
\beq
T(k) = \left (1 + \frac{A k^2}{\log{(1 + Bk)}}\right )^{-1}~.
\label{eq:transfer}
\eeq
Equations (\ref{eq:pow_spec}) \& (\ref{eq:transfer}) illustrate
the `turn around' of the power spectrum from
its primordial scale invariant form $P(k) \propto k$ on the largest scales to 
$P(k) \propto k^{-3}\log^2k$ on
small scales. 
(The precise location of the turn-around and the amplitude of $P(k)$ depend upon
specific details of the cosmological model, see for instance \cite{bbks86}.) 

\begin{figure}[ht]
\centering
\includegraphics[width=10cm]{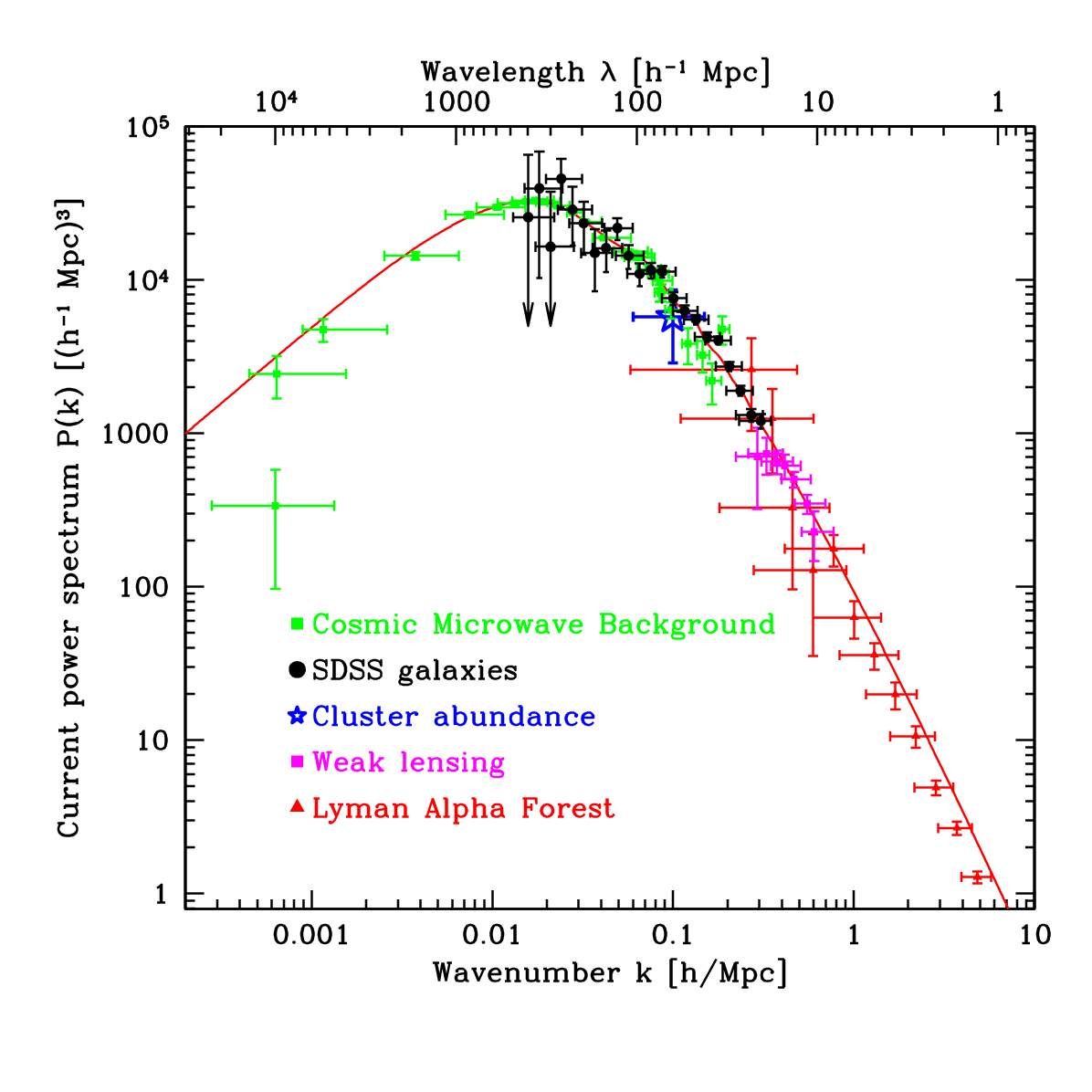}
\caption{\footnotesize
The power spectrum inferred from observations of large scale structure, the
Lyman$\alpha$ forest, gravitational lensing and the CMB.
The solid line shows the power spectrum prediction for a flat scale-invariant
LCDM model with $\Omega_m = 0.28$,
$\Omega_b/\Omega_m = 0.16$, 
$h = 0.72$; from Tegmark et al. ~\protect\cite{tegmark03a}.
}
\label{fig:power}
\end{figure}

The `standard' cold dark matter (SCDM) paradigm, which assumed that
$\Omega_{\rm CDM} = 1$,
was introduced during the early 1980's at roughly the same time when HDM was perceived
to be in trouble (see \cite{kolbt90,khlopov,coles,sc95} for 
references to earlier work on 
this subject).
Although SCDM was very successful in explaining a host of
observational details, it was clear already a decade ago, that the processed
power spectrum of SCDM lacked sufficient power on large scales and so fell short of explaining
the angular two point correlation function for galaxies
on scales $\sim 50$ Mpc
\cite{efstath}.
The relevant cosmological quantity in this case is the shape of the power spectrum of density
perturbations, which for CDM-like models, can be characterised by the `shape parameter'
$\Gamma = \Omega_m h$. SCDM models with $\Omega_m = 1$ and the HST-determined
value $h \simeq 0.7$
predict $\Gamma \simeq 0.5$ which is much larger than the observed value $\Gamma = 0.207
\pm 0.030$
inferred from observations of galaxy clustering in the sloan digital sky survey (SDSS)
\cite{pope}.
A modification of SCDM called LCDM assumes that, in addition to CDM the universe consists of
a smoothly distributed component called a cosmological constant or a Lambda-term.
LCDM models with $h \simeq 0.7$ and $\Omega_m = 0.3$ predict a smaller value for
the shape parameter, $\Gamma \simeq 0.2$, and the resulting amplitude and shape of the
power spectrum is  in excellent
agreement with several different sets of observations as demonstrated in figure
\ref{fig:power}.

From (\ref{eq:pow1}), (\ref{eq:pow_spec}) \& (\ref{eq:transfer}) we find that 
on small scales, the contribution to the
{\em rms} density fluctuation
from a given logarithmic interval in $k$ is
\beq
\left (\frac{\delta\rho}{\rho}\right )^2_k \sim k^3 P_f(k) \propto \log^2{k}~,
\eeq
which illustrates the fact that, although the smallest scales are the first to go non-linear,
there is significant power to drive gravitational instability rapidly 
to larger scales in this model.
Indeed, detailed N-body simulations of large scale structure show that filaments defining
the cosmic web first form on the smallest scales. 
The scale-length characterizing the cosmic web grows
as the universe expands, until at the present epoch the cosmic web consists of a 
fully developed supercluster-void network with a scale-length of several tens of Megaparsec
\cite{sss96,sss03,masaar,rien}. 

Promising candidates for cold dark matter include a 
$100 - 1000$ GeV particle called a neutralino.
The neutralino is a weakly interacting massive particle
(WIMP). As its name suggests it is neutral and is 
a fermionic partner to the gauge and Higgs bosons (usually called the `bino, wino and
higgsino'). It is believed that the lightest supersymmetric particle will be stable due
to R-parity which makes the neutralino an excellent candidate for cold dark matter
(see \cite{roszko99,jungman96} for reviews of particle dark matter).
A radically different particle candidate for cold dark matter is an ultra-light 
pseudo-Goldstone boson called an axion with a mass of only $m_a \sim 10^{-5 \pm 1}$ eV. 
Although ultralight, the axion is `cold' because it was created as a zero-momentum condensate.
Its existence is a by-product of an attempt to resolve QCD of what is commonly called
the `strong CP problem' which arises because non-perturbative effects in QCD give rise to
an electric dipole moment for the neutron -- in marked contrast with observations
\cite{kolbt90}.
Other candidates for non-baryonic cold dark matter include string theory motivated
modulii fields \cite{brustein}; non-thermally produced super-heavy particles having a mass
$\sim 10^{14}$ GeV and dubbed Wimpzillas \cite{kolb}; as well as axino's and gravitino's
-- superpartners of the axion and graviton respectively \cite{roszko99}. 

Since WIMP's cluster gravitationally, one should expect to find a flux of these particles
in our own solar system and attempts are being made to determine dark matter particles
by measuring the scattering of WIMP's on target nucleii through nuclear recoils.
Now the earth orbits the sun with a velocity $\simeq 30$ km/sec, even as
the sun orbits the galaxy with $v_{M_\odot} \simeq 220$ km/sec.
Furthermore the plane of the Earth's orbit is inclined at an angle of $60^\circ$ to the
glactic plane, because of which the dark matter flux on Earth is expected to be larger in June
(when the Earth's velocity and the Sun's velocity add together) than in December
(when these two velocities subtract).
The resulting rate variation is about $7\%$ between the flux measured
during summer and winter. Precisely such a signal was reported by
the DAMA experiment whose data (collected since 1996) appears to show a yearly modulation
with greater events reported in June than in December \cite{dama}.
However results obtained by the DAMA group remain controversial since they
have not been substantiated by other groups
which report negative results for similar searches
(see \cite{munoz,khalil} for recent reviews on this subject).

\begin{figure}[ht]
\centering
\includegraphics[width=5cm]{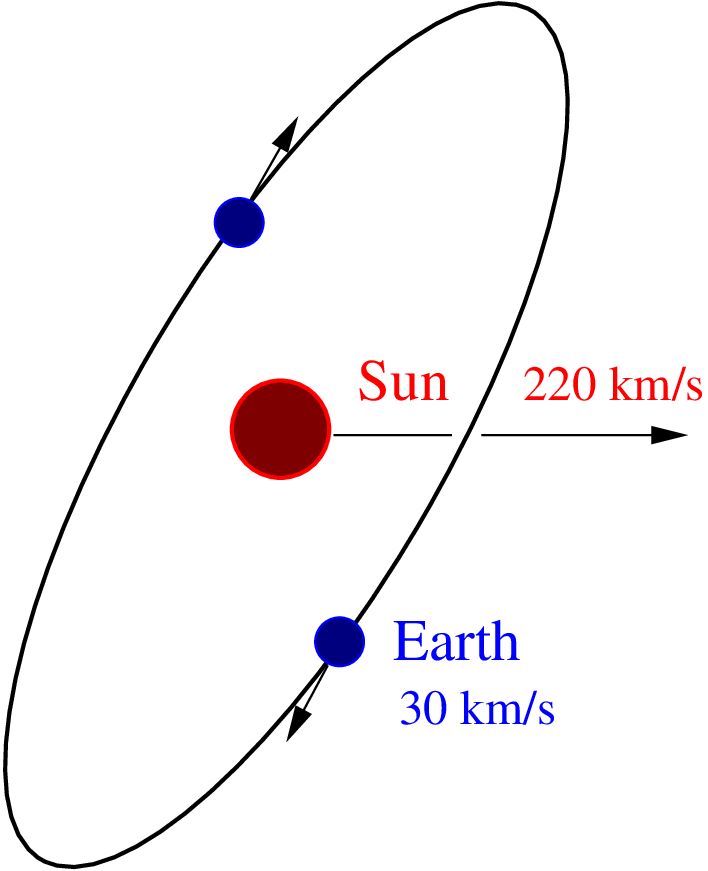}
\caption{\footnotesize
The Earth's motion around the Sun; from Khalil and Munoz (2001).
}
\bigskip
\medskip
\label{fig:sol}
\end{figure}

Despite the excellent agreement of LCDM with large scale observations, some concerns
have recently been expressed about the ability of this model to account for a number
of smaller scale observations which can be summarized as follows:

\begin{itemize}

\item {\em The substructure problem} is used to describe the fact that the cold dark matter
model (with or without a cosmological constant) predicts an excessive number of dark matter
subhaloes (or substructure) within a larger halo. If one (perhaps naively) associates each halo with a 
gravitationally bound baryonic object then the predicted number of dwarf-galaxy satellites
within the local group 
exceeds the observed number by over an order of magnitude. 
Indeed, detailed N-body simulations as well as theoretical estimates predict around
1000 dark matter satellites in our local group which is much larger than the
40 or so observed at present 
\cite{klypin99,moore99,kauffmann93,bullock00, somerville02,benson02,mateo98,tasi02,evans01}.

\item {\em The cuspy core problem} 

CDM predicts a {\em universal density profile}
for dark matter halos in the wide range $10^7 M_\odot - 10^{15} M_\odot$ which
applies both to galaxy clusters as well as individual galaxies including dwarfs and 
LSB's. \footnote{Low Surface Brightness Galaxies (LSB's) are dominated by their dark matter
content and therefore provide particularly good astrophysical objects with which to
test dark matter models.}
The density profile originally suggested by Navarro, Frenk and White \cite{nfw} is
\beq
\rho(r) = \rho_0\left (r_s/r\right )\bigg\lbrack 1 + \left (\frac{r}{r_s}\right 
)\bigg \rbrack^{-2}~,
\eeq
which gives $\rho \propto r^{-1}$ for $r \ll r_s$ and 
$\rho \propto r^{-3}$ for $r \gg r_s$, where $r_s$ is the scale radius
and $\rho_0$ is the characteristic halo density. (Other groups using higher resolution
computations found somewhat steeper
density profiles at small radii, such as $\rho \propto r^{-1.5}$ \cite{moore99b,jing00}.)

{\em The cuspy core problem} refers to the apparent contradiction between
N-body experiments -- which show that the density profile in CDM halos
has a $1/r$ (or steeper) density cusp at the center,
and observations -- which appear to favour significantly
 shallower density cores in galaxy clusters as well as in individual dwarf and LSB galaxies 
 (see
 \cite{flores94,burkert95,bolatto02,blok97,primack01,tasi02,klypin04,rhee03,sand03,ma04} for
 detailed discussions of this issue).

\end{itemize}

Although disconcerting,
given the very considerable success of LCDM in explaining gravitational clustering
on large scales, 
it may at this point be premature to condemn this model on the basis of small scale observations
alone. It could be that the difficulties alluded to above are a result of an oversimplification
of the complex physical processes involved and that a more
careful analysis of the baryonic physics on small scales including the hydrodynamical effects of
star formation and supernova feedback needs to be undertaken.
For instance both dwarfs and LSB's have very shallow potential wells, a strong burst
of star formation and supernova activity may therefore empty dark matter halos
of their baryonic content resulting in a large number of `failed galaxies' and
providing a possible resolution to the `satellite catastrophe'. 
(The failed galaxies will act as gravitational lenses and should therefore
be detectable through careful observations.)
Other explanations
include the effects of tidal stripping recently discussed in \cite{klypin04}.
Likewise issues involving beam smearing, the influence of bars and the interaction
of baryons and dark matter in the central regions of galaxies and clusters
could be intricately linked with the central cusp issue and must be better understood 
if one wishes to seriously test the CDM hypothesis on small scales.

In concluding this discussion on dark matter I would like to briefly mention 
Modified Newtonian Dynamics (MOND) which, in some circles, is regarded as 
an alternative 
to the dark matter hypothesis. 
 As the name suggests, MOND is a modification of Newtonian physics
 which proposes to explain the flat rotation curves of galaxies without invoking
 any assumptions about dark matter. Briefly, MOND assumes that Newtons law of inertia
 ($F = ma$) is modified at sufficiently low accelerations ($a < a_0$) to
 \beq
 {\bf F} = m{\bf a} \mu(a/a_0)~,
 \eeq
where $\mu(x) = x$ when $x \ll 1$ and $\mu(x) = 1$ when $x \gg 1$ \cite{milgrom,sanders02}.
It is easy to see that this results in the modification of the conventional formula
for gravitational acceleration ${\bf F} = m{\bf g_N}$, resulting in the following
relation between the true acceleration and the Newtonian value: $a = \sqrt{g_N a_0}$.
For a body orbiting a point mass $M$, $g_N = GM/r^2$. Since the centripetal acceleration
$a = v^2/r$ now equals the {\em true} acceleration $a$, one gets
\beq
v^4 = GMa_0~,
\eeq
\ie for sufficiently low values of the acceleration
the rotation curve of an isolated body of mass $M$ does not depend upon the radial
distance $r$ at which the velocity is measured, in other words not only does
this theory predict 
flat rotation curves it also suggests that the individual halo associated with a galaxy is
infinite in extent ! 
(This latter prediction may be a problem for MOND since recent galaxy-galaxy
lensing results \cite{hoekstra02}
suggest that galaxy halo's may have a maximum extent of about 0.5 Mpc.)
The value of $a_0$ needed to explain observations is $a_0 \sim 10^{-8}$cm/s$^2$ 
which is of the same order as $cH_0$ ! This has led supporters of this hypothesis
to conjecture that MOND may reflect ``the effect of cosmology on local particle
dynamics'' \cite{sanders02}.
Although MOND gives results which are in good agreement with observations of individual 
galaxies, it is not clear whether it is as successful for explaining clusters
for which strong gravitational lensing indicates a larger mass concentration at cluster
centers than accounted for by MOND \cite{sanders02,combes02}.
Another difficulty with MOND is that it is problematic to
embed this theory within a more comprehensive
relativistic theory of gravity and hence, at present, it is not clear what predictions
a MOND-type theory may make for gravitational lensing and other curved 
space-time effects. For some recent developments in this direction see
\cite{beken04}.


To summarise, current observations make a strong case for clustered, non-baryonic 
dark matter 
to account for 
as much as a third of the total matter density in the Universe
$\Omega_m \simeq 1/3$. The remaining two-thirds is thought to reside in a 
relative smooth component having large negative pressure and called 
Dark Energy. 

\section{Dark Energy}
\index{dark energy}

\subsection{The cosmological constant and Vacuum energy}
\index{Vacuum energy and the cosmological constant}

Type Ia
supernovae, when treated as standardized candles, suggest that the expansion
of the universe is speeding up rather than slowing down. The case for an
accelerating universe also receives independent support from CMB and large 
scale structure studies.
All three data sets can be simultaneously satified if one postulates that the dominant
component of the universe is relatively smooth, has a large negative 
pressure and $\Omega_{\rm DE} \simeq 2/3$.

The simplest example of dark energy is a cosmological constant, introduced by Einstein
in 1917.
The Einstein equations, in the presence of the cosmological constant, aquire the form
\beq
R_{ik} - \frac{1}{2} g_{ik} R = \frac{8\pi G}{c^4} T_{ik} + \Lambda g_{ik}~.
\label{eq:einstein}
\eeq
Although Einstein originally introduced the cosmological constant ($\la$) into
the left hand side of his field equations, it has now become conventional
to move the $\la$-term to the RHS, treating it as an effective form of matter.
In a homogeneous and isotropic Friedmann-Robertson-Walker (FRW) universe
consisting of pressureless dust (dark matter) and $\la$,
the Raychaudhury equation, which follows from (\ref{eq:einstein}), takes the form
\beq
{\ddot a} = -\frac{4\pi G}{3}a \rho_m + \frac{\la}{3}\,a~.
\label{eq:einstein1}
\eeq
Equation (\ref{eq:einstein1}) can be rewritten in the form of a force law\footnote{Relation 
(\ref{eq:einstein2}) can also be derived from the Schwarzschild-de Sitter metric describing a spherically
symmetric matter 
distribution of mass $M$ embedded in de Sitter space with the cosmological constant
$\Lambda$:
\beq
ds^2 = \left ( 1+\frac{2\varphi}{c^2}\right )c^2dt^2 - 
\left ( 1+\frac{2\varphi}{c^2}\right )^{-1}dR^2
- R^2\left (d\theta^2 + \sin^2{\theta}d\phi^2 \right )
\eeq
where 
\beq
\varphi = -\frac{GM}{R} - \frac{\Lambda}{6}R^2
\eeq
is the gravitational potential and the associated force is
${\cal F} = -\partial\varphi/\partial R$.}:
\beq
{\cal F} = -\frac{GM}{R^2} + \frac{\la}{3}R, ~~~(R \equiv a)
\label{eq:einstein2}
\eeq
which demonstrates that the cosmological constant gives rise to a
{\em repulsive force} whose value increases with distance. 
The repulsive nature of $\la$ could be responsible for the acceleration of
the universe as demonstrated in (\ref{eq:einstein1}).

Although introduced into physics in 1917, the physical basis for a cosmological
constant remained a bit of a mystery until the 1960's, when it was realised that
zero-point vacuum fluctuations must respect Lorenz invariance and therefore
have the form $\tik = \la g_{ik}$ \cite{zel68}.
As it turns out, the vacuum expectation value of the energy momentum is 
divergent both for bosonic and fermionic fields, and this gives rise to
what is known as `the cosmological constant problem'.
Indeed the effective cosmological constant generated by vacuum fluctuations is
\beq
\frac{\la}{8\pi G} = \langle T_{00}\rangle_{\rm vac}
\propto \int_0^\infty \sqrt{k^2 + m^2}k^2 dk~,
\eeq
since the integral diverges as $k^4$ one gets an infinite value for the
vacuum energy. Even if one chooses to `regularize' $\tik$ by imposing an
ultraviolet cutoff at the Planck scale, 
one is still left with an enormously large value for
the vacuum energy $\langle T_{00}\rangle_{\rm vac}
\simeq c^5/G^2\hbar \sim 10^{76}$GeV$^4$
which is 123 orders of magnitude larger than
the currently observed $\rho_\la \simeq 10^{-47}$GeV$^4$.
A smaller ultraviolet cut-off does not fare much better since
a cutoff at the QCD scale results in 
$\Lambda_{QCD}^4 \sim 10^{-3}$GeV$^4$, which is still forty orders of magnitude larger than observed.

In the 1970's the discovery of supersymmetry led to the hope that,
since bosons and fermions
(of identical mass) contribute equally but with opposite sign to the
vacuum expectation value of physical quantities,
the cosmological constant problem may be resolved by a judicious
balance between bosons and fermions in nature. 
However supersymmetry (if it exists) is broken at the low temperatures
prevailing in the universe today and on this account one should
expect the cosmological
constant to vanish in the early universe, but to reappear during late
times when the temperature has dropped below $T_{\rm SUSY}$.
This is clearly an undesirable scenario and almost the very opposite
of what one is looking for, since, a large value of $\la$ at an
early time is useful from the viewpoint of inflation, whereas a very
small current value of $\la$ is in agreement with observations \cite{ss00,sahni02}.

In the absence of a resolution to the cosmological constant problem
the following possibility connecting
the vacuum energy with the SUSY and Planck scales may be worth exploring
\cite{ss00,sahni02}. 
The mass scale
associated with the scale of supersymmetry breaking
in some models,
$M_{\rm SUSY} \sim 1$ TeV,
lies midway between the Planck scale and $10^{-3}$ eV.
One could conjecture that the
small observed value of the cosmological constant
$\rho_\la \simeq (10^{-3} eV)^4$
is associated with the vacuum in a theory which had a
fundamental mass scale $M_X \simeq M_{\rm SUSY}^2/M_{\rm Pl}$, such that
$\rho_{\rm vac} \sim M_X^4 \sim (10^{-3} eV)^4$.

The cosmological constant is also relevant from the perspective of models
with spontaneous symmetry breaking \cite{wein89}.
Indeed, if one examines the Lagrangian
\ber
{\cal L} &=& \half ~g^{ij}\partial_i\phi\partial_j\phi - V(\phi),\nonumber\\
~V(\phi) &=& V_0 -\half \mu^2\phi^2 + \frac{1}{4}\lambda\phi^4,
\label{eq:inf2a}
\eer
one notices that the symmetric state at $\phi = 0$ is unstable and the system
settles in the ground state $\phi = +\sigma$
or $\phi = -\sigma$, where $\sigma = \sqrt{\mu^2/\lambda}$, thereby breaking the
reflection symmetry $\phi \leftrightarrow - \phi$ present in the Lagrangian.
For $V_0 = 0$ this potential gives rise to  
a large negative cosmological constant
$\la_{eff} = V(\phi = \sigma) = - \mu^4/4\lambda$ in the broken symmetry state.
This embarassing situation can be avoided only if one chooses a value for $V_0$ which 
almost exactly cancels $\la_{eff}$, 
namely $V_0 \simeq + \mu^4/4\lambda$ so that
$\la_{eff}/8\pi G = V_0 - \mu^4/4\lambda \simeq 10^{-47}$GeV $^4$.

The cosmological consequences of this rather ad-hoc `regularization'
exercise are instructive.
Unless the value of $\la_{eff}$ lies in a very small window, the universe will
be a very different place from the one we are used to.
For instance a negative value of the $\la$-term 
$\la_{eff}/8\pi G < - 10^{-43}$GeV $^4$ will cause the universe to
recollapse (the effect of $\la$ is attractive now instead of being repulsive)
less than a billion years after the big bang -- a period which is 
much too short for galaxies to form and for life
(as we know it) to emerge.
On the other hand a large positive $\la_{eff}/8\pi G > 10^{-43}$GeV$^4$
makes the universe accelerate much before the present epoch,
thereby inhibiting structure formation and precluding the emergence of life.

The very small window in $\la$ which allows life to emerge has led 
some cosmologists to propose 
anthropic arguments for the existence of
a small cosmological constant \cite{barrow86,wein98,vilenk00,wein01}.
One such possibility is the following
``if our big bang is just one of many big bangs, with a wide range of
vacuum energies, then it is natural that some of these big bangs should have
a vacuum energy in the narrow range where galaxies can form, and of course it is
 just these big bangs in which there could be astronomers and
 physicists wondering about the vacuum energy'' \cite{wein01}.

I will not discuss the anthropic argument any further in these lectures but
will point the interested reader to \cite{wein98,vilenk00,wein01} for
further discussion of this issue.

It is important to note that there is no known
fundamental symmetry in nature which will set the
value of $\la$ to zero. In its absence, the small observed value of the dark energy
remains somewhat of a dilemma which remains to be fully understood and resolved.
\footnote{The important role played by symmetries is illustrated
by the U(1) gauge symmetry of electrodynamics whose presence 
implies a zero rest mass for the photon.
No analogous symmetry exists for the neutrino and recent experiments do indicate that
neutrino's could have a small mass.}

\begin{figure}[ht]
\centering
\includegraphics[width=7cm]{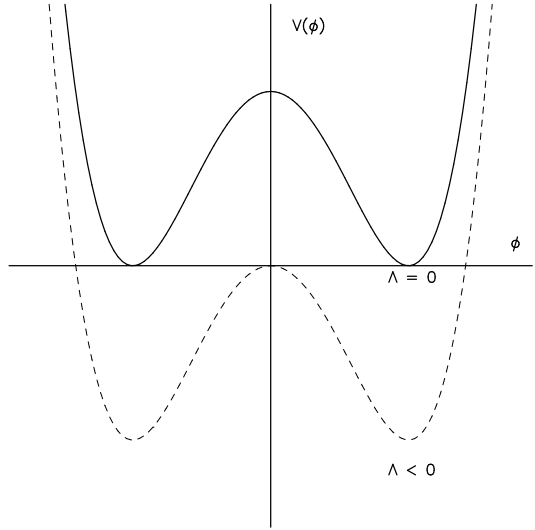}
\caption{\footnotesize
Spontaneous symetry breaking in many field theory models takes the form of
the Mexican top hat potential shown above.
The dashed line shows the potential before the cosmological constant has been
`renormalized' and the solid line after.
(From Sahni and Starobinsky 2000.)
}
\bigskip
\medskip
\label{fig:ssb}
\end{figure}

\subsection{Dynamical models of Dark Energy}

The cosmological constant is but one example of a form of matter (dark energy)
which could drive an accelerated phase in the history of our universe.
Indeed, (\ref{eq:einstein1}) is easily generalised to 
\beq
\frac{\ddot a}{a} = -\frac{4\pi G}{3} \sum_i(\rho_i + 3p_i) = 
-\frac{4\pi G}{3} \sum_i\rho_i(1+3w_i)~,
\label{eq:einstein3}
\eeq
where the summation is over all
forms of matter present in the universe with equation of state $w_i = p_i/\rho_i$.
Eqn. (\ref{eq:einstein3}) together with its companion equation
\begin{equation}
H^2 \equiv \left (\frac{\dot a}{a}\right )^2 = \frac{8\pi G}{3} \sum_i
\rho_i - \frac{k}{a^2}
\label{eq:hubble1}
\end{equation}
completely describes the dynamics of a FRW universe ($k/a^2$ is the Gaussian
curvature of space).

Clearly a universe consisting of only a single component will accelerate if
$w < -1/3$.
Fluids satisfying $\rho + 3p \geq 0$ or $w \geq -1/3$ are said to satisfy the
`strong energy condition' (SEC). We therefore find
that, in order to accelerate, `dark energy' must {\rm violate the SEC}.
Another condition which is usually assumed to be sacrosanct, but has recently
been called into question is the `weak energy condition' (WEC) 
$\rho + p \geq 0$ or $w \geq -1$. Failure to satisfy the WEC can result in
faster-than-exponential expansion for the universe
and in a cosmic `Big Rip', which we shall come to in
a moment.

It is often more convenient to rewrite (\ref{eq:einstein3}) in terms of 
the `deceleration parameter'
\beq
q = - \frac{\ddot a}{a H^2} = \sum_i (\frac{4\pi G\rho_i}{3 H^2})(1 + 3w_i)
= \frac{\left ( 1 + 3 w_{\rm X} \Omega_{\rm X}\right )}{2}~,
\label{eq:decel}
\eeq
where $\Omega_i = 8\pi G\rho_i/3H^2$ and 
we have assumed a flat universe with $\Omega_{\rm m} + \Omega_{\rm X} = 1$
($\Omega_{\rm X} \equiv \Omega_{\rm DE}$).
The condition for accelerated expansion ($q < 0$) is equivalent to
\beq
w_X < - \frac{1}{3(1-\Omega_{\rm m})}~,
\eeq
which leads to 
\ber
w &<& - \frac{1}{3} ~~ {\rm for}~ \Omega_m = 0~,\label{eq:half}\\
w &<& - \frac{1}{2} ~~ {\rm for}~ \Omega_m = 1/3 ~.
\label{eq:third}
\eer

Eqn. (\ref{eq:hubble1}) can be used to develop an expression for the 
Hubble parameter $H \equiv\ {\dot a}/a$
in terms of the cosmological redshift
$z = a_0/a(t) - 1$:
\beq
H(z) = H_0 \left\lbrack \Omega_m (1+z)^3 +
\Omega_X (1+z)^{3(1+w)} \right\rbrack^{1/2},
\label{eq:hubble2}
\eeq
where $H_0 = H(z=0)$  is the present value of the Hubble parameter,
$\Omega_m = 8\pi G\rho_{0m}/3H_0^2$,
$\Omega_X = 8\pi G\rho_{0{\rm DE}}/3H_0^2$, describe the dimensionless density
of matter and dark energy respectively, 
($w \equiv w_{\rm DE}$), and we have
made the assumption of a flat universe so that 
$\Omega_m + \Omega_X = 1$.

In LCDM cosmology $w=-1$, $\Omega_\Lambda = \la/3H_0^2$,
and the expansion factor has the elegant form
\cite{ss00}
\beq
a(t) \propto \left(\sinh {{3\over 2}}
\sqrt{{\Lambda\over 3}}ct\right)^{2/3}~,
\label{eq:interpol}
\eeq
which smoothly interpolates between a matter dominated universe in the
past ($a \propto t^{2/3}$) and accelerated expansion in the
future ($a \propto \exp{\sqrt{\frac{\Lambda}{3}}t}$).

We are now in a position to appreciate the evidence for an accelerating
universe which originates in observations of the light flux from
high redshift type Ia
supernovae. Type Ia supernovae are extremely bright objects, 
($M_B \simeq -19.5$) which makes them ideally suited for studying
the properties of the universe at large distances.

The light flux received from a distant
supernova is related to its absolute luminosity ${\cal L}$ and its `luminosity distance' 
$d_L$ 
through the relation
\begin{equation}
F = \frac{\cal L}{4\pi d_L^2}.
\end{equation}
If one views this problem from within the Newtonian perspective then,
since the geometry of space is Euclidean, 
$d_L = \sqrt{x^2+y^2+z^2}$. In general relativity,
on the other hand, the
geometry of space can be non-Euclidean, and the luminosity distance to
an object located at redshift $z$ will, in general, depend both upon
the geometry of space as well as the expansion history of the universe.
Indeed, it can be shown that in a spatially flat and
expanding FRW universe, the luminosity distance
has the form
\beq
d_L(z) = (1+z) \int_0^z\frac{dz'}{H(z')}.
\label{eq:lum}
\eeq

The luminosity distance is shown in Fig. ~1 for a number of
cosmological models with varying amounts of $\Omega_m$ \& $\Omega_\Lambda$.
The limiting case $\Omega_m = 1$, $\Omega_\Lambda = 0$ corresponds to
standard cold dark matter 
 (SCDM) in which the universe decelerates as a
 weak power law
$ a(t) \propto t^{2/3}$.
The other extreme example $\Omega_\Lambda = 1$, $\Omega_m = 0$
describes the de Sitter universe (also known as steady state cosmology)
which accelerates at the steady rate
$ a(t) \propto \exp{\sqrt{\frac{\Lambda}{3}}t}$.
From figure ~1 we see that a supernova at
redshift $z = 3$ will appear 9 times brighter in SCDM 
than it will in de Sitter space !

\begin{figure}[ht]
\centering
\includegraphics[width=7cm]{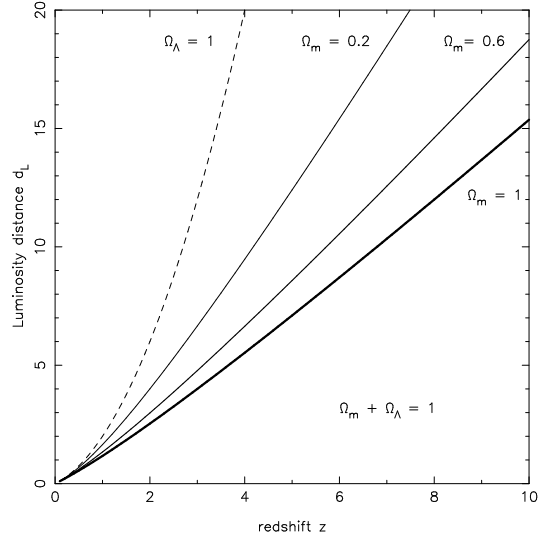}
\caption{\footnotesize
The luminosity distance $d_L$ (in units of $H_0^{-1}$)
is shown as a function of 
cosmological redshift
$z$ for spatially
flat cosmological models with $\Omega_m +
\Omega_\Lambda = 1$. 
Heavier lines correspond to larger values of $\Omega_m$.
The dashed line shows the luminosity distance in the spatially flat de Sitter
universe ($\Omega_\Lambda = 1$). From Sahni and Starobinsky \cite{ss00}.
}
\bigskip
\medskip
\label{fig:lum_dis}
\end{figure}

\begin{figure}[ht]
\centering
\includegraphics[width=10cm]{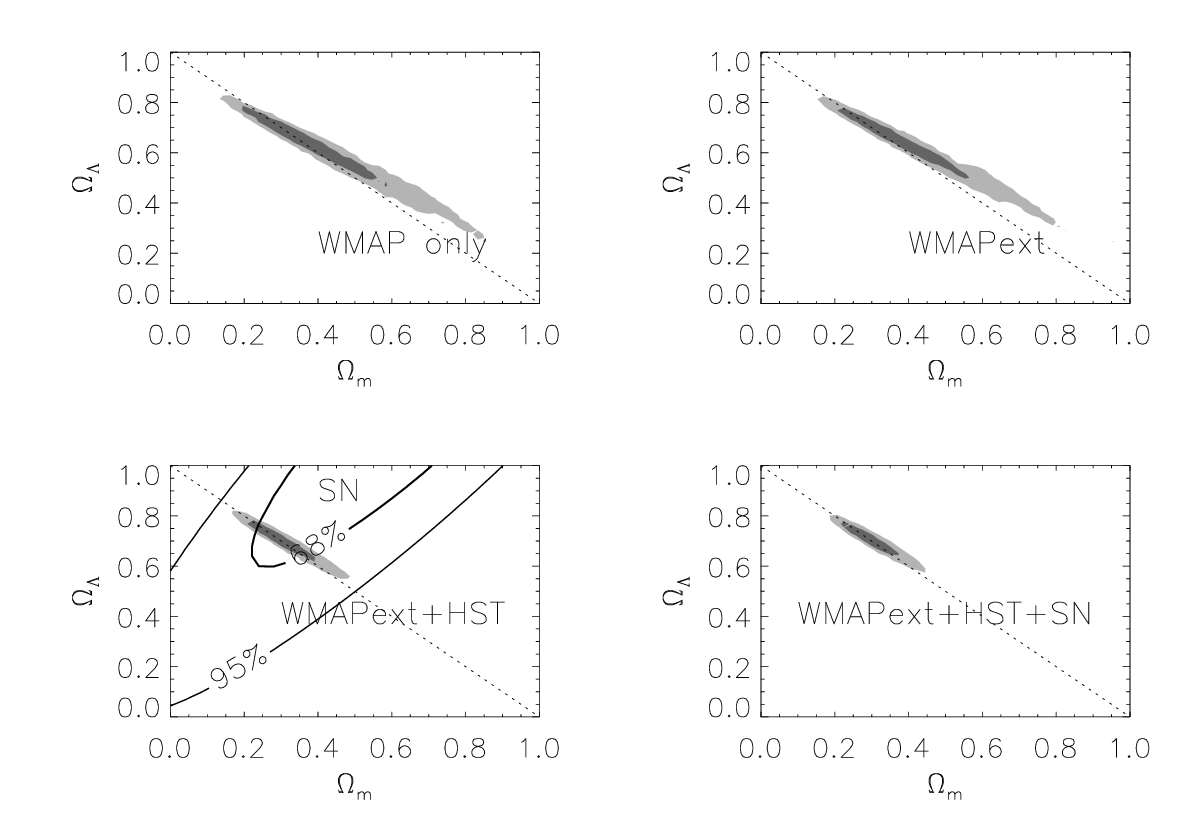}
\caption{\footnotesize
Constraints on the density of dark matter $\Omega_m$ and dark energy 
in the form of a cosmological constant $\Omega_\la$, determined using 
WMAP (upper left), WMAP + other CMB experiments (WMAPext; upper right), 
WMAPext + HST key project data (lower left) and WMAPext + HST + supernova data
(lower right); from Spergel et al (2003).
}
\label{fig:wmap}
\end{figure}

Systematic studies of type Ia supernovae have revealed that:

\begin{itemize}

\item Type Ia SNe are excellent stadardized candles.
The dispersion in peak supernova luminosity is small: 
$\Delta m \simeq 0.3$, 
and the corresponding change in intensity is about $25\%$.
In addition the light curve of a type Ia supernova is correlated with its
peak luminosity \cite{phillips93} to a precision of $\sim 7\%$, so that
{\em brighter supernovae take longer to fade.}
(Type Ia SNe take roughly 20 days to rise from relative obscurity to maximum light.)
This allows us to `standardize' supernova light curves thereby reducing the scatter
in their luminosities to $\sim 12\%$ which turns type Ia supernovae into
very good standard
candles.

\item Type Ia supernovae at higher redshifts are consistently dimmer than
their counterparts at lower redshifts relative to what might be expected in
SCDM cosmology. If type Ia 
supernovae are treated as standard candles
then, assuming systematic effects such as cosmological evolution and dimming by 
intergalactic
dust are either not vitally important or have been corrected for, 
the systematic dimming of high$-z$ SNe can be
interpreted as evidence for an accelerated expansion of the universe caused
by a form of `dark energy' having large negative pressure.

\end{itemize}
The evidence for an accelerating universe from high redshift type Ia supernovae
has now received independent support from an analysis of CMB fluctuations 
together with the HST key project determination of the Hubble parameter.
Interestingly, the degeneracy in parameter space 
$\lbrace \Omega_m, \Omega_\la\rbrace$ arising from SNe observations
is almost orthogonal to the degeneracy which arises from CMB measurements.
This principle of `cosmic complementarity' serves to significantly reduce 
the errors on $\Omega_m$ \& $\Omega_\la$ when the two sets of observations
are combined, as shown in figure \ref{fig:wmap}.

If dark energy is described by an unevolving equation of state
$w = p_X/\rho_X$, then the transition between
deceleration and acceleration (${\ddot a} = 0$)
occurs at the redshift 
\beq
(1 + z_{\rm a})^{-3w} = -(1 + 3w)\frac{\Omega_X}{\Omega_m} ~ w < 0.
\label{eq:decel1}
\eeq
Another important redshift describes the epoch when the densities in dark matter
and dark energy are equal
\beq
(1 + z_{\rm eq})^{3w} = \left (\frac{\Omega_m}{\Omega_X}\right )~.
\label{eq:equal}
\eeq
Substituting $\Omega_\Lambda = 0.7, \Omega_m = 0.3$ we find
$z_{\rm a} \simeq 0.73$, $z_{\rm eq} \simeq 0.37$ for LCDM.
The fact that the acceleration of the universe is a fairly recent phenomenon
illustrates the `cosmic coincidence' puzzle according to which we appear
to live during a special epoch when the densities in dark energy and in
dark matter are almost equal.
A recent origin for the acceleration epoch is supported by
supernova observations which suggest a decelerating universe at
$z \ggeq 0.5$ \cite{riess04}. It is important to note that
dark enegy models with an unevolving equation of state need to have their
initial conditions properly `tuned' in order to dominate the universe at 
precisely the present epoch. This problem is most acute for the cosmological
constant.
Since the cosmological constant does not evolve 
while both matter and radiation evolve rapidly 
($\rho_m \propto a^{-3}$, $\rho_r \propto a^{-4}$),
it follows that the small current value 
$\rho_\Lambda = \la/8\pi G \simeq 10^{-47}$ GeV$^4$
implies 
$\rho_\Lambda/\rho_r \simeq 10^{-123}$ at the Planck time (when the
temperature of the universe was $T \sim 10^{19}$ GeV),
or $\rho_\Lambda/\rho_r \simeq 10^{-55}$ at the time of the electroweak
phase transition ($T \sim 100$ GeV).
Thus an extreme fine-tuning of initial conditions is required
in order to ensure that $\rho_\Lambda/\rho_m \sim 1$ today !

The fine tuning problem which plagues $\la$ also affects DE models in which
$w = {\rm constant} \neq -1$. A combined analaysis of CMB, galaxy clustering and 
supernovae data indicates that a constant equation of 
state for dark energy must satisfy $w < -0.82$ at the 95\% confidence level
\cite{spergel03,tegmark03b}, and it is easy to show that for these models
the fine tuning (and cosmic coincidence) problems are almost as acute as
they are for the cosmological constant.
This constraint on $w$ also virtually rules out two interesting DE candidates
based on topological defect models: a tangled network of cosmic strings 
$w \simeq -1/3$ and domain walls $w \simeq -2/3$.

\subsection{Quintessence}

It is interesting that the fine tuning problem facing dark energy models
with a constant equation of state can be alleviated
if we assume that the equation of state is time dependent.
An important class of models having this property are scalar fields (quintessence)\footnote{
Quintessence is named after the all pervasive fifth element 
of ancient philosophical thought. Note that the quintessence Lagrangian is
the same as that used for Inflationary model building.}
which couple minimally to gravity so that their Lagrangian density and 
energy momentum tensor is
\beq
{\cal L} = \half\dot{\phi}^2 - V(\phi)~,
\label{eq:quint_lagr}
\eeq

\beq
\rho \equiv T_0^0 = \half\dot{\phi}^2 + V(\phi),
~p \equiv -T_\alpha^\alpha = \half\dot{\phi}^2 - V(\phi)~,
\label{eq:scalar_em}
\eeq
where we have assumed, for simplicity, that the field is homogeneous.
Potentials which are sufficiently steep to satisfy
$\Gamma \equiv V''V/(V')^2 \geq 1$ have the interesting property that
scalar fields rolling down such a potential
approach a common evolutionary path from a wide range of initial
conditions \cite{track}. In these so-called `tracker' models
the scalar field density (and its equation of state) remains close
to that of the dominant background matter during most of cosmological evolution.

\begin{figure}[ht]
\centering
\includegraphics[width=8cm]{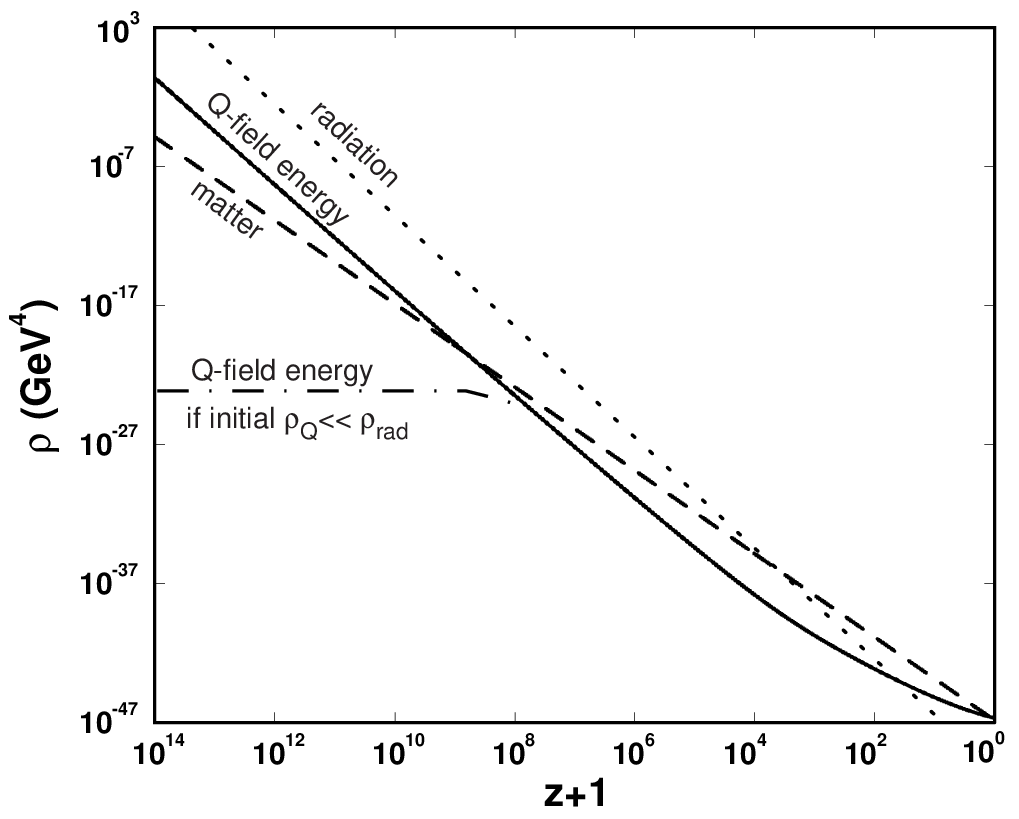}
\caption{\footnotesize
The quintessence Q-field while rolling an inverse power law potential
tracks first radiation then matter, before coming to dominate
the energy density of the universe at present.
If the initial value of the Q-field density is small then $\rho_{\rm Q}$ remains
constant until $\rho_{\rm Q} \sim \rho_{\rm rad}$, and then
follows the tracker trajectory. From Zlatev, Wang and Steinhardt ~\protect\cite{track}.
}
\bigskip
\medskip
\label{fig:quint}
\end{figure}

An excellent example of a tracker potential 
is  provided by $V(\phi) = V_0/\phi^{\alpha}$ \cite{ratra}. During tracking
the ratio of the energy density of the scalar field (quintessence)
to that of radiation/matter gradually increases
$\rho_\phi/\rho_B \propto t^{4/(2 + \alpha)}$
while its equation of state remains marginally smaller than the
background value $w_\phi = (\alpha w_B - 2)/(\alpha + 2)$.
For large values of $\phi$ this potential becomes flat ensuring
that the scalar field rolls sufficiently slowly ($\dot{\phi}^2 \ll V(\phi)$) 
to allow the universe to accelerate.
Note that for quintessence fields the conditions (\ref{eq:half}) \&
(\ref{eq:third}) translate into 
\ber
w_\phi < -\frac{1}{3} ~&\Rightarrow& ~\dot{\phi}^2 < V(\phi)~,\nonumber\\
w_\phi < -\frac{1}{2} ~&\Rightarrow& ~ \dot{\phi}^2 < \frac{2}{3} V(\phi)~.
\eer
(Current observations imply $\alpha < 2$.)

An extreme example of quintessence is provided by the exponential potential
$V(\phi) = V_0\exp{(-\sqrt{8\pi}\lambda\phi/M_p)}$ \cite{ratra,wetterich88},
where $M_p = 1/\sqrt{G}$ is the Planck mass.
In this case
\beq
\frac{\rho_\phi}{\rho_{B} + \rho_\phi} = \frac{3(1 + w_B)}{\lambda^2}
= {\rm constant} < 0.2,
\label{eq:exp}
\eeq
$\rho_{B}$ is the background energy density while $w_B$ is the
associated background equation of state.
The lower limit $\rho_\phi/\rho_{\rm total} < 0.2$ arises because of
nucleosynthesis constraints which prevent the energy density in
quintessence from being large initially (at $t \sim few $  sec.).
Equation (\ref{eq:exp})
suggests that the exponential potential will remain subdominant if
it was so initially.
An interesting potential which interpolates between an exponential
and a power law can however give rise to late time acceleration from tracker-like
initial conditions \cite{sw00}
\beq
V(\phi) = V_0[\cosh{\lambda\phi} - 1]^p,
\label{eq:cosh}
\eeq
has the property that $w_\phi \simeq w_B$ at early times
whereas $\langle w_\phi\rangle = (p - 1)/(p + 1)$ at late times.
Consequently (\ref{eq:cosh})
describes {\em quintessence} for $p \leq 1/2$ and pressureless
`cold' dark matter (CDM) for 
$p = 1$.
Thus the cosine hyperbolic potential (\ref{eq:cosh}) is able to describe both
dark matter and dark energy within a tracker framework (also see \cite{lopez,arbey03}).

Remarkably, quintessence can even accomodate a constant equation of
state ($w = {\rm constant}$) by means of the potential \cite{ss00,sahni03,um00}
\beq
V(\phi)\propto \sinh ^{{2(1+w)\over  w}}(C\phi+D)~,
\eeq
with suitably chosen values of $C, D$.


\bigskip
\begin{table*}[tbh!]
\begin{center}
\begin{minipage}[h]{0.9\linewidth} \mbox{} \vskip 18pt
\begin{tabular}{lll}
\hline
Quintessence Potential & ~~~~Reference\\\hline
& \\
$V_0\exp{(-\lambda\phi)}$ & Ratra \& Peebles (1988), Wetterich (1988), \\
& Ferreira \& Joyce (1998)\\
& \\
$m^2\phi^2, \lambda\phi^4$ &  Frieman et al (1995)\\
& \\
$V_0/\phi^\alpha, \alpha > 0$ &  Ratra \& Peebles (1988) \\
& \\
$V_0\exp{(\lambda\phi^2)}/\phi^\alpha $ & Brax \& Martin (1999,2000)\\
& \\
$V_0(\cosh{\lambda\phi} - 1)^p$ & Sahni \& Wang (2000)\\
& \\
$V_0 \sinh^{-\alpha}{(\lambda\phi)}$ & Sahni \& Starobinsky (2000),
Ure\~{n}a-L\'{o}pez \& Matos (2000)\\
& \\
$V_0(e^{\alpha\kappa\phi} + e^{\beta\kappa\phi})$ & Barreiro, Copeland \& Nunes (
2000)\\
& \\
$V_0(\exp{M_p/\phi} - 1)$ & Zlatev, Wang \& Steinhardt (1999)\\
& \\
$V_0[(\phi - B)^\alpha + A]e^{-\lambda\phi}$ & Albrecht \& Skordis (2000)\\

& \\
\hline
\end{tabular}
\caption{}
\end{minipage}
\end{center}
\end{table*}

Quintessence models can be divided into two categories: 
models which roll to large values of
$\phi/m_P \ggeq 1$ and models for which $\phi/m_P \ll 1$ at the present epoch.
An important concern for the former is the effect of quantum corrections
which, if large, could alter the shape of the quintessence potential \cite{kolda98,burgess03,doran02,bassett03}.
An important related issue is that the coupling between standard model fields
and quintessence must be small in order to have evaded detection.
Moreover even small couplings between quintessence and
standard model fields can give rise to interesting
changes in cosmology as shown in
\cite{amendola,maccio}.

I would like to end this section by mentioning that, 
due to the shortage of time I have not been able to cover all of the DE models
suggested in the literature (a number that is growing rapidly !) 
For this reason these lectures will not discuss DE due to vacuum
polarization\cite{sh98,parker99}, k-essence \cite{mukhanov}, 
Cardassian expansion \cite{ktf}, Quasi-steady state cosmology \cite{jvn02},
scalar-tensor models \cite{amendola99,bertolami99,chiba99,pbm99,uzan99,perrotta99a,ritis99,beps00,perrotta03}. For other interesting approaches see
\cite{axenides04,hsu04a,hsu04b,elizalde03,nojiri03c,gu03,li01,li02,mongan01,shapiro99,zhu01}.
A partial list of some popular quintessence models is given in Table 1,
and the reader is also referred to the dark energy reviews in
\cite{ss00,carroll01,pr02,sahni02,paddy03}.

\subsection{Dark energy in braneworld models}
\label{sec:brane}

Inspired by the Randall-Sundrum \cite{rs} scenario, braneworld cosmology
suggests that we could be living on a three dimensional `brane'
which is embedded in a higher (usually four) dimensional bulk.
According to such a scheme, all matter fields are confined to the brane
whereas the graviton is free to propogate in the brane as well as in the bulk
(see the lectures by Roy Maartens in this volume and \cite{maartens03}
for a comprehensive discussion
of Braneworld cosmology.) Within the RS setting the
 equation of motion of a scalar field propogating on the brane is
 \beq
 {\ddot \phi} + 3H {\dot \phi} + V'(\phi) = 0.
 \label{eq:kg}
 \eeq
 where \cite{brane}
\ber
H^2 &=& \frac{8\pi}{3 m^2}\rho (1 + \frac{\rho}{2\sigma}) + 
\frac{\Lambda_4}{3} +\frac{\cal E}{a^4}~,\nonumber\\
\rho &=& \half{\dot\phi}^2 + V(\phi)~.
\label{eq:brane}
\eer
${\cal E}$ is an integration constant which transmits bulk graviton
influence onto the brane. The brane tension
$\sigma$ provides a
relationship between the four dimensional Planck mass ($m$) and the
five-dimensional Planck mass ($M$)
\beq
m = \sqrt{\frac{3}{4\pi}}\bigg (\frac{M^3}{\sqrt{\sigma}}\bigg ) ~.
\eeq
$\sigma$ also relates the four-dimensional cosmological constant $\Lambda_4$
on the brane to the five-dimensional (bulk) cosmological constant 
$\Lambda_{\rm b}$ through
\beq
\Lambda_4 = \frac{4\pi}{M^3}\bigg (\Lambda_{\rm b} + \frac{4\pi}{3M^3}\sigma^2
\bigg). 
\eeq

\begin{figure}[ht]
\centering
\includegraphics[width=9cm]{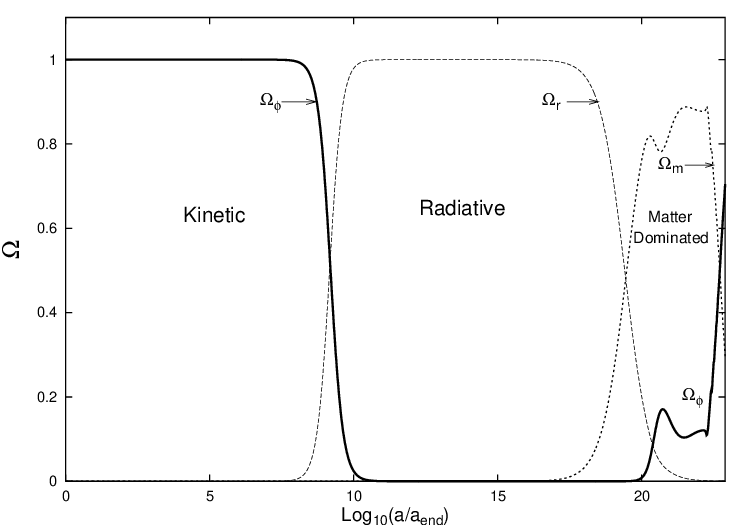}
\caption{\footnotesize
The post-inflationary  density parameter $\Omega$ is plotted for
the scalar field (solid line) radiation (dashed line) and cold dark
matter (dotted line)
in the quintessential-inflationary model
decribed by (\ref{eq:cosh}) with $p = 0.2$.
Late time oscillations of the scalar field ensure that
the mean equation of state turns negative
$\langle w_\phi\rangle \simeq -2/3$, giving rise to the current
epoch of cosmic acceleration with $a(t) \propto t^2$ and present day values
$\Omega_{0\phi} \simeq 0.7, \Omega_{0m} \simeq 0.3$.
From Sahni, Sami and Souradeep ~\protect\cite{sami}.
}
\bigskip
\medskip
\label{fig:quint_inf}
\end{figure}

 Note that (\ref{eq:brane}) contains an
 additional term $\rho^2/\sigma$ whose presence can be attributed to
junction conditions imposed at the bulk-brane boundary. Because of
this term 
the damping experienced by the scalar field as it rolls down its
potential {\em dramatically increases} so that inflation can be sourced by potentials
which are normally too steep to produce slow-roll.
Indeed the slow-roll parameters
in braneworld models
(for $V/\sigma \gg 1$) are \cite{maartens}
\beq
\epsilon \simeq 4\epsilon_{\rm FRW} (V/\sigma)^{-1},\,
\eta \simeq 2\eta_{\rm FRW} (V/\sigma)^{-1},\,
\label{eq:slow-roll}
\eeq
illustrating that slow-roll ($\epsilon, \eta \ll 1$) is easier to achieve when
$V/\sigma \gg 1$. 
Inflation can therefore arise for the very steep potentials associated with
quintessence such as
$V \propto e^{-\lambda\phi}$, $V \propto \phi^{-\alpha}$ etc.
This gives rise to the intriguing 
possibility that both inflation and quintessence may be sourced
by one and the same scalar field. Termed  `quintessential inflation',
these models have been examined in 
\cite{peebles99,copeland,lidsey,sami,majumdar,curvaton1,curvaton2,dimo04,bi1,sami1,sami2}. 
An example of quintessential inflation is shown in figure \ref{fig:quint_inf}.

A radically different way of making the Universe accelerate 
was suggested in \cite{DDG,ss02}.
The braneworld model developed by Deffayet, Dvali and Gabadadze (DDG) was radically
different from the RS model in that both the bulk cosmological constant and the brane
tension were set to zero, while a curvature term was introduced 
in the brane action so that the theory was described by
\beq
S = M^3 \int_{\rm bulk} {\cal R} + m^2 \int_{\rm brane}  R 
+ \int_{\rm brane} L_{\rm matter}  \, .
\eeq
The rationale for the $\int_{\rm brane} R$ term is that quantum effects
associated with
matter fields are likely to give rise to such a term in the Einstein action
as discussed by Sakharov in his development of induced gravity 
\cite{sakharov67}.

The resulting Hubble parameter in the DDG braneworld is
\beq\label{eq:ddg}
H = \sqrt{\frac{8\pi G \rho_{\rm m}}{3} + \frac{1}{l_c^2}} +
\frac{1}{l_c}~,
\eeq
where $l_c = m^2/M^3$ is a new length scale determined by the four dimensional Planck mass 
$m$ and and the five dimensional Planck mass $M$ respectively.
An important property of this model is that the 
acceleration of the universe is not caused by the
presence of any `dark energy'. Instead, 
since gravity becomes five dimensional on length scales $R >
l_c = 2H_0^{-1}(1-\Omega_{\rm m})^{-1}$, one finds that the expansion of the
universe is modified during {\em late times} instead of early times as in the RS model.

A more general class of
braneworld models which includes RS cosmology and the DDG brane as subclasses
was developed in \cite{CH,Shtanov1} and is described by the action
\begin{equation} \label{action}
S = M^3 \int_{\rm bulk} \left( {\cal R} - 2 \Lambda_{\rm b}
\right) + \int_{\rm brane} \left( m^2 R - 2
\sigma \right) + \int_{\rm brane} L_{\rm matter} \, .
\end{equation}
For $\sigma = \la_{\rm b} = 0$ (\ref{action})
reduces to the action describing the DDG model, whereas for $m = 0$ it 
describes the Randall-Sundrum model.

As demonstrated by Sahni and Shtanov \cite{ss02} the braneworld which follows 
from the action (\ref{action}) describes an accelerating universe at late times with
the Hubble parameter 
\ber \label{eq:hubble_brane}
{H^2(z) \over H_0^2} &=& \Omega_{\rm m} (1\!+\!z)^3 + \Omega_\sigma +2 \Omega_l \mp 
\nonumber\\
&&2 \sqrt{\Omega_l}\,
\sqrt{\Omega_{\rm m} (1\!+\!z )^3 + \Omega_\sigma + \Omega_l +
\Omega_{\la_{\rm b}}} \, ,
\eer
where
\beq \label{eq:omegas}
\Omega_l = {1 \over l_c^2 H_0^2}~, ~\Omega_{\rm m} =  {\rho_{0m} \over 3 m^2 H_0^2}~, ~\Omega_\sigma
= {\sigma \over 3 m^2H_0^2}~,
~\Omega_{\la_{\rm b}} = - {\la_{\rm b} \over 6 H_0^2}~.
\eeq
(The $\mp$ signs refer to the two different ways in which the brane can be
embedded in the bulk, both signs give rise to interesting cosmology \cite{ss02}.)
As in the DDG model
$l_c \sim H_0^{-1}$ if $M \sim 100$ MeV.  On short length
scales $r \ll l_c$ and at early times, one recovers general
relativity, whereas on large length scales $r \gg l_c$ and at late
times brane-related effects begin to play an important role.  
Indeed by setting $M=0$ ($\Omega_l = 0$)
(\ref{eq:hubble_brane}) reduces to the LCDM model
\beq
{H^2(z) \over H_0^2} = \Omega_{\rm m} (1\!+\!z)^3 + \Omega_\sigma~,
\eeq
whereas for $\sigma = \la_{\rm b} = 0$ 
(\ref{eq:hubble_brane}) reduces to the DDG braneworld. 
An important feature of the braneworld (\ref{eq:hubble_brane}) 
is that it can lead to an effective equation of state of dark energy 
$w_{\rm eff} \leq -1$. This is easy to see from the expression for the
current value of the effective equation 
of state \cite{ss02}
\begin{equation}
w_0 = {2 q_0 - 1 \over 3 \left( 1 - \Omega_{\rm m} \right)} = - 1 \pm
{\Omega_{\rm m} \over 1 - \Omega_{\rm m}} \, {\sqrt{\Omega_\ell \over
\Omega_{\rm m} + \Omega_\sigma + \Omega_\ell + \Omega_{\Lambda_{\rm b}}}} \, ,
\label{eq:brane_state}
\end{equation}
we find that $w_0 < -1$ when we take 
the lower sign in (\ref{eq:brane_state}), which corresponds to choosing one of 
two possible embeddings of this braneworld in the higher dimensional bulk.
(The second choice of embedding gives $w_0 > -1$.)

It is also possible, in this model, for the acceleration of
the universe to be a transient phenomenon which ends once the universe returns
to matter
dominated expansion after the current accelerating phase.
As discussed in \cite{ss02} such a braneworld will not have an event horizon
and may therefore help in reconciling an accelerating universe with the demands of
string/M-theory.
Other possibilities of obtaining dark energy from extra dimensions 
have been discussed in 
\cite{albrecht02,peri02,chen00,pietroni03,peloso03,maroto04,burgess04,nojiri03d}.
The possibility that DE could arise 
due to modifications of gravitational physics has also been 
examined in \cite{starkman03,capozziello03,carroll03b,dolgov03,mofatt04,nojiri03a,nojiri03b}.

\subsection{Chaplygin gas}

A completely different route to dark energy is provided by the Chaplygin gas
\cite{chap1}
which obeys the equation of state
\beq
p_c = - A/\rho_c~.
\label{eq:chap_state}
\eeq
The conservation equation $d E = -p dV \Rightarrow d(\rho a^3) = -pd(a^3)$
immediately gives
\beq
\rho_c=\sqrt{ A+\frac{B}{a^6}} = \sqrt{ A+B (1+z)^6}\,\,,
\eeq
where $B$ is a constant of integration.
Thus
the Chaplygin gas behaves
like pressureless dust at early times and like a cosmological constant
during very late times. 

The Hubble parameter for a universe containing cold dark matter and
the Chaplygin gas is given by
\beq\label{eq:hub_chap}
H(z) = H_0\left\lbrack \Omega_m(1+z)^3 + \frac{\Omega_m}{\kappa}\sqrt{\frac{A}{B} +
(1+z)^6}\right\rbrack^{1/2}\,\,,
\eeq
where $\kappa = \rho_{0m}/\sqrt{B}$ and
it is easy to see from (\ref{eq:hub_chap}) that
\beq\label{eq:chap_def}
\kappa = \frac{\rho_{0m}}{\rho_c}(z \to \infty)\,\,.
\eeq
Thus, $\kappa$ defines the ratio between CDM and the Chaplygin gas
energy densities at the commencement of the matter-dominated stage. It
is easy to show that
\beq
A = B \left\lbrace \kappa^2 \left( \frac{1-\Omega_m}{\Omega_m} \right)^2 - 1
\right\rbrace \,\,.
\eeq
It is interesting that the Chaplygin gas can be derived from an underlying Lagrangian
in two distinct ways:

\begin{itemize}

\item One can derive it from a quintessence Lagrangian (\ref{eq:quint_lagr})
with the 
potential \cite{chap1}
\beq
V(\phi) = \frac{\sqrt{A}}{2}\left (\cosh{3\phi} + \frac{1}{\cosh{3\phi}}
\right )~.
\eeq

\item The Chaplygin gas can also be derived from the 
Born-Infeld form of the Lagrangian density

\beq
{\cal L} = -V_0\sqrt{1-\phi_{,\mu}\phi^{,\mu}}~,
\label{eq:born}
\eeq
where $\phi_{,\mu} \equiv \partial{\phi}/\partial x^\mu$.
For time-like $\phi_{,\mu}$ one can define a four velocity
\beq
u^\mu = \frac{\phi^{,\mu}}{\sqrt{\phi_{,\alpha}\phi^{,\alpha}}}~,
\eeq
this leads to the standard form for the hydrodynamical energy-momentum tensor
\beq
T_{\mu \nu} = (\rho + p)u_\mu u_\nu - pg_{\mu \nu}~,
\eeq
where \cite{fks02}
\beq
\rho = \frac{V_0}{\sqrt{1-\phi_{,\mu}\phi^{,\mu}}}~,~~
p = - V_0\sqrt{1-\phi_{,\mu}\phi^{,\mu}}~,
\eeq
\ie we have recovered (\ref{eq:chap_state}) with $A = V_0^2$.

\end{itemize}

The fact that the properties of the Chaplygin gas interpolate between
those of CDM and a $\la$-term led to the hope that the CG might
provide a conceptual framework for a unified model of dark matter and
dark energy. It should however be noted that
in contrast to CDM and baryons, the sound
velocity in the Chaplygin gas $v_c= \sqrt{dp_c / d\rho_c}={\sqrt
A / \rho_c}$ quickly grows $\propto t^2$ during the
matter-dominated regime and becomes of the order of the velocity of
light at present (it approaches light velocity asymptotically in the
distant future ).  Thus, when one examines classical inhomogeneities,
the properties of the Chaplygin gas during the matter-dominated epoch
appear to be rather unusual and resemble those of hot dark matter 
rather than CDM,
despite the fact that the Chaplygin gas formally
carries negative pressure \cite{alam03a}.

A `generalized Chaplygin gas' has also been proposed for which
$p \propto -1/\rho^\alpha$. The equation of state in this case is
\beq
w(a) = - \frac{|w_0|}{\big\lbrack |w_0| + \frac{1-|w_0|}{a^{3(1+\alpha)}}\big\rbrack}~,
\eeq
which interpolates between $w = 0$ at early times ($a \ll 1$) and 
$w = -1$ at late times ($a \gg 1$); $w_0$ is the current equation of state 
at $a = 1$. (The constant $\alpha$ regulates the transition time in the
equation of state.)
WMAP, supernovae and large scale sructure data have all been used to test 
Chaplygin gas models; see \cite{chap2,chap3,chap4,chap5,chap6,chap7,chap8,chap9,chap10,chap11,chap12}.

\subsection{Is Dark Energy a Phantom ?}
\label{sec:phantom}

In an influential paper Caldwell \cite{caldwell} noticed that a very good fit to
the supernova-derived luminosity distance was provided by
dark energy which violated the weak energy condition so that
$w < -1$. He called this Phantom dark energy.\footnote{Phantom takes its name
from Part I of
the Star Wars movie series -- the Phantom Menace.}
Indeed, a study of high-z SNe \cite{knop03} finds that the DE equation of state 
has a $99\%$ probability of being $< -1$ if no priors
are placed on $\Omega_m$ ! 
When these SNe results are combined with CMB and 2dFGRS
the 95\% confidence limits on an unevolving equation of state are
$-1.61 < w < -0.78$ \cite{knop03}, which is consistent with estimates made
by other groups \cite{spergel03,tegmark03b}.

A universe filled with Phantom energy has some interesting but bizarre properties.

\begin{itemize}

\item If $t_{\rm eq}$ marks the epoch when the densities in matter and phantom
energy are equal then the
expansion factor of a universe dominated by phantom energy grows as
\begin{equation}
a(t) \simeq a\left( t_{\rm eq} \right) \left[(1 + w) \frac{t}{t_{\rm eq}} - w
\right]^{2 \big/ 3(1+w)} \, , ~~ w < -1 ~,
\end{equation}
and therefore {\em diverges in a finite\/} amount
of cosmic time
\begin{equation}
a(t) \to \infty \ \ \mbox{as} \ \ t \to t_{\rm BR} = \left( \frac{w}{1+w} \right)
t_{\rm eq} \, .
\end{equation}
By substituting $w < -1$ into (\ref{eq:hubble2}) we immediately find that the
Hubble parameter also diverges as $t \to t_{\rm BR}$, implying that an infinitely
rapid expansion rate for the universe has been reached in a {\em finite\/}
time. The divergence of the Hubble parameter is associated with
the divergence of phantom density which grows without bound
\begin{equation} \rho(t) \propto \left[(1 + w) \frac{t}{t_{\rm eq}} - w
\right]^{-2} , \end{equation} and reaches a singular value in a finite interval of
time $\rho(t) \to \infty$, $t \to t_{\rm BR}$. Thus a universe dominated by Phantom energy
culminates in a future curvature singularity (`Big Rip') at which the
notion of a classical space-time breaks down.
(See also \cite{star00,chiba00,caldwell,innes,caldwell03,carroll03,frampton03a,frampton03b, 
singh,johri03,alcaniz03,bridle03}.)

\item The ultra-negative phantom equation of state suggests that the effective velocity
of sound in the medium $v = \sqrt{|dp/d\rho|}$ can become larger than
the velocity of light in this model.

\item Although a dynamical model of phantom energy can be constructed with the `wrong' sign
of the kinetic term, see (\ref{eq:scalar_em}), such models 
are plagued with instabilities at the quantum level \cite{cline} 
which makes their existence suspect.

It should be pointed out that phantom is not the only way to 
get $w < -1$. A model with similar properties but sharing none of 
phantom's pathologies is the braneworld model of \cite{ss02,as02}, which has
$w_{\rm eff} < -1$ today but does not run into a `Big Rip' in the future.

\end{itemize}

\subsection{Reconstructing Dark Energy and the Statefinder diagnostic}

In view of the considerable number of dark energy models suggested in the literature, 
it becomes meaningful to ask whether we can reconstruct
the properties of DE from observations in a model independent manner.
This indeed may be possible if one notices that the
the Hubble parameter is related to the luminosity distance \cite{star98,saini00}
\beq
H(z) = \left[ \frac{d}{dz} \left( \frac{d_L(z)}{1+z} \right) \right]^{-1},
\eeq
and that, in the case of quintessence, the scalar field potential
as well as its equation of state can be directly expressed in terms
of the Hubble parameter and its derivative \cite{star98,saini00}
\ber
{8\pi G\over 3H_0^2} V(x)\ &=& {H^2\over H_0^2}
-{x\over 6H_0^2}{dH^2\over dx} -{1\over 2}\Omega_m\,x^3,
\label{eqn:Vzed}\\
{8\pi G\over 3H_0^2}\left({d\phi\over dx}\right)^2 &=&
{2\over 3H_0^2 x}{d\ln H\over dx}
-{\Omega_m x\over H^2}, ~ x = 1+z,\\
&&\\
w_\phi (x) \equiv {p\over\varepsilon} &=&
\frac{(2x/3) d\ln H/dx -1}{1-\left(H^2_0/H^2\right)
\Omega_m x^3}.
\eer
Both the quintessence potential $V(\phi)$ as well as
the equation of state $w_\phi (z)$ may therefore be reconstructed provided
the luminosity  distance $d_L(z)$ is
known to reasonable accuracy from observations.

\begin{figure}[ht]
\centering
\includegraphics[width=6cm]{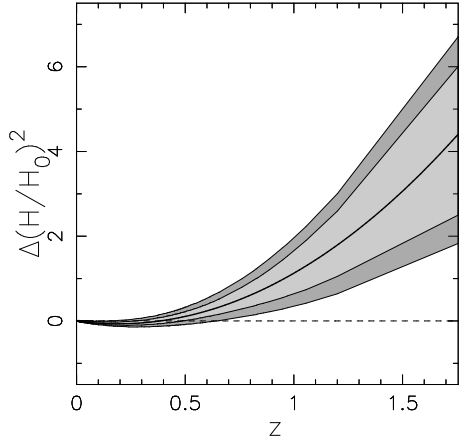}
\caption{\footnotesize
The relative difference between the Hubble parameter reconstructed from SNe data
and the LCDM value is shown as a function of redshift. SNe data from Tonry et al (2003)
were used for the reconstruction.
The best-fit is represented by the thick solid line assuming $\Omega_m = 0.3$.
The light (dark) grey
contours represents the $1\sigma$ ($2\sigma$) confidence levels
around the best-fit.
The dashed horizontal line shows LCDM. From Alam, Sahni, Saini and Starobinsky
~\protect\cite{alam03b}.
}
\bigskip
\medskip
\label{fig:sn_hubble}
\end{figure}

In practice it is useful to have an ansatz for either one of three
cosmological quantities:
$d_L(z)$, $H(z)$ or $w(z)$, which can then be used for cosmological reconstruction
\cite{saini00,recon1,recon2,recon3}.
Popular fitting functions discussed in the literature include:

\n (i) An ansatz for the dark energy \cite{sahni03}
\beq
\rho_{\rm DE}(x) =  \sum_{i=0}^N A_i x^i, ~~ x = 1+z~.
\label{eq:taylor}
\eeq

\n (ii) Fitting functions to the dark energy equation of state
\cite{albrecht,linder}:
\ber 
w(z) &=& \sum_{i=0}^N w_iz^i\,\,,\nonumber\\ 
w(z) &=& w_0+\frac{w_1 z}{1+z}~.
\label{eq:wpol}
\eer
The fitting parameters $w_i$, $A_i$ are obtained by matching
to observations. In practice the first few terms in either series
(\ref{eq:taylor}), (\ref{eq:wpol}) is sufficient since the current SNe data 
are quite noisy;
see \cite{chiba,albrecht,copeland03,efstathiou,maor02,alam03a} 
for a discussion of these issues.
An example of cosmological reconstruction of the Hubble parameter from 
SNe data is shown in figure (\ref{fig:sn_hubble}); see also \cite{wangm,leandros}.

The SNe inventory is increasing dramatically every year and so are increasingly precise
measurements of galaxy clustering and the CMB. To keep pace with 
the better quality observational data which will soon become available
and the increasing
sophistication of theoretical modelling,
a new 
diagnostic of DE called `Statefinder' was introduced in \cite{sahni03}.

\begin{figure}[ht]
\centering
\includegraphics[width=8cm]{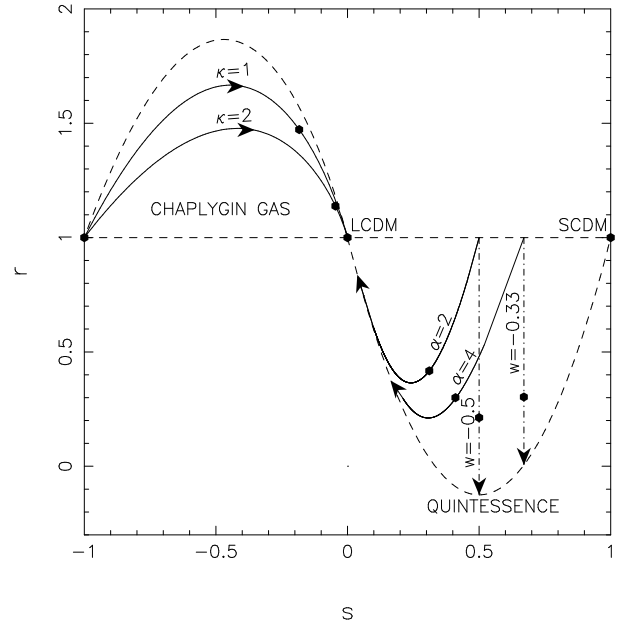}
\caption{\footnotesize
The time evolution of the statefinder pair
$\lbrace r,s \rbrace$ for quintessence models and the Chaplygin gas.
Solid lines to the right of LCDM represent tracker potentials
$V=V_0/\phi^{\alpha}$, while those to the left correspond to the Chaplygin gas.
Dot-dashed lines represent DE with a
constant equation of state $w$.
Tracker models tend to approach the LCDM fixed
point ($r=1, s=0$) from the right at $t \to \infty$,  whereas the Chaplygin gas
approaches LCDM from the left.
For Chaplygin gas
$\kappa$ is the ratio between matter density and the
density of the Chaplygin gas at early times.
The dashed curve in the lower right is the envelope of all
quintessence models, while the dashed curve in the upper left is the
envelope of Chaplygin gas models (the latter is described by $\kappa =
\Omega_m/1-\Omega_m$).  The region outside the dashed curves is forbidden for
both classes of dark energy models.
The ability of the Statefinder to differentiate between dark energy models
is clearly demonstrated.
From Alam, Sahni, Saini and Starobinsky
~\protect\cite{alam03a}.
}
\label{fig:statefinder}
\end{figure}

The statefinder probes the expansion dynamics of the
universe through higher derivatives of the expansion factor 
$\atridot$ and is a natural companion to the deceleration parameter 
which depends upon ${\ddot a}$
(\ref{eq:decel}). The statefinder pair $\statei$ is defined as follows:
\ber\label{eq:statefinder1}
r &\equiv& \frac{\atridot}{a H^3} = 1 + \frac{9w}{2} \omx(1+w) -
\frac{3}{2} \omx \frac{\dot{w}}{H} \,\,,\\
s &\equiv& \frac{r-1}{3(q-1/2)} = 1+w-\frac{1}{3} \frac{\dot{w}}{wH}\,\,.
\eer
Inclusion of
the statefinder pair $\statei$, increases the number of cosmological
parameters to four\footnote{$r$ has also been called
`cosmic jerk' in \cite{visser03}}: $H$, $q$ $r$, $s$.
The Statefinder is a `geometrical' diagnostic in the sense that it depends upon the 
expansion factor and hence upon the metric describing space-time.
An important property of the Statefinder is that spatially
flat LCDM corresponds to the fixed point
\beq
\statei\bigg\vert_{\rm LCDM} = \lbrace 1,0 \rbrace ~. 
\eeq

Departure of a given DE model from this fixed point provides a good way of
establishing the `distance' of this model from LCDM \cite{alam03a}.
As demonstrated in \cite{sahni03,alam03a,chap4,zimdahl03} 
the Statefinder can successfully differentiate 
between a wide variety of DE models including the cosmological constant,
quintessence, the Chaplygin gas,
braneworld models and interacting DE models; an example is provided in
figure \ref{fig:statefinder}.

\subsection{Big Rip, Big Crunch or Big Horizon ? --
The fate of the Universe in dark energy models}
\label{sec:future}
The nature of dark energy affects the future of our Universe in a very significant way.
If DE is simply the cosmological constant, then the universe will accelerate for ever.
%
Of great importance is the fact that an accelerating LCDM universe develops an
event horizon similar to the one surrounding a black hole \cite{star00}. 
Consider an event ($r_1,t_1$) which we wish to observe
at our location at $r = 0$. Setting $ds^2 = 0$ we get
\beq
\int_0^{r_1}\frac{dr}{\sqrt{1-\kappa r^2}} = \int_{t_1}^{t}\frac{cdt'}{a(t')}.
\label{eq:future1}
\eeq
Any event in the universe
will one day be observed by us if the integral in the RHS of
(\ref{eq:future1}) {\em diverges}
as $t \to \infty$. For power law expansion this clearly implies
$a \propto t^p$, $p < 1$, \ie a {\em decelerating universe}.
In an accelerating universe exactly the opposite is true, the integral in the
RHS converges signalling the presence of an event horizon.
In this case our civilization will receive signals only from those events
which satisfy \cite{ss00} 
\beq
\int_0^{r_1}\frac{dr}{\sqrt{1-\kappa r^2}} \leq
\int_{t_1}^{\infty}\frac{cdt'}{a(t')}.
\label{eq:future2}
\eeq
For de Sitter-like expansion
$a = a_1\exp{H(t - t_1)}, H = \sqrt{\la/3}$, we get $r_1 = c/a_1H$, so that the
proper distance to the event horizon is $R_H = a_1r_1 = c/H$.
In LCDM cosmology, 
\beq
H \equiv H(t \to \infty) = \sqrt{\la/3} = H_0\sqrt{1 - \Omega_m}~,
\eeq
and the proper distance to the horizon is 
\beq
R_H = \frac{c}{H_0\sqrt{1 - \Omega_m}} \simeq 3.67 h^{-1}~{\rm Gpc}~,
\eeq
if $\Omega_m \simeq 1/3$. Thus our observable universe will progressively shrink
as astrophysical bodies which are not gravitationally bound to the local group 
get pushed to distances beyond $R_H$. (More generally,
horizons exist in a universe which begins to perpetually accelerate
after a given point of time \cite{hellerman01,fischler01,sahni02}.
To this category belong models of dark energy
with equation of state $-1 < w < -1/3$, as well as `runaway
scalar fields' \cite{stein01} which satisfy
$V,V',V'' \to 0$ and $V'/V, V''/V \to 0$ as
$\phi \to \infty$.)

The presence of an event horizon implies that, at any given moment of time
$t_0$, there is a `sphere of influence' around our civilization. This sphere has an
associated redshift $z_H$, and a celestial body having $z > z_H$ will be unreachable
by any signal emitted by our civilization now or in the future; 
$z_H \simeq 1.8$ in LCDM cosmology with $\Omega_\la \simeq 2\Omega_m \simeq 2/3$. 
Thus all celestial bodies with $z > 1.8$ lie beyond our event horizon and there is
no possibility of causal contact with any of them. 

Interestingly, an N-body simulation tracking the future of an LCDM universe has shown that
$\sim 100$ billion years from now the observable universe will consist of only
a single massive galaxy within our event horizon
-- the merger product of the Milky Way and Andromeda galaxies \cite{loeb03}.
Furthermore, since the growth of large scale structure freezes in an
accelerating universe, the mass distribution of bound objects will cease to
evolve after about $30$ billion years.

This somewhat gloomy future scenario is not absolutely
essential and can be avoided if the currently observed acceleration of the universe
is a transient phenomenon.\footnote{Accelerating cosmologies without
event horizons are important in a different context.
Since the conventional S-matrix approach may not work in a universe with an
event horizon, models with horizons 
may pose a serious challenge to a fundamental theory of
interactions such as string/M-theory.} 
Just such a possibility exists in a class of braneworld models \cite{ss02} in which the 
current accelerating phase is succeeded by a decelerating matter dominated regime. 
Quintessence potentials can also have this property, as discussed in \cite{barrow00}.
An interesting class of transiently accelerating DE models is constructed around a
scalar field potential
which decays with time and becomes negative at late times 
\cite{frieman95,choi,ngwilt,kl02,kl03,ass03}.
An example is $V = V_0 \cos{\phi/f}$ which describes axionic quintessence
\cite{frieman95,choi,ngwilt,ass03}. Such a universe recollapses in the future 
when $H(t_0 + \Delta T) = 0$, and contracts thereafter towards a 
`Big Crunch' singularity. 
Supernova observations indicate that, for typical decaying potentials, the universe
will not collapse for at least $\Delta T \simeq 20$ Gyrs \cite{ass03}.

DE models have also been proposed which encounter
a `quiescent singularity' while expanding. At the `quiescent singularity' 
the second derivative of the 
expansion factor diverges while its first derivative remains finite \cite{ss02b,kamen03}
(\ie ${\ddot a} \to -\infty$, 
${\dot a} \simeq {\rm constant}$).
In such models the expansion of the universe 
`brakes' to a virtual standstill 
as the universe approaches the singular regime at
which invariants of the space-time metric diverge 
($R_{iklm}R^{iklm} \to \infty$) while, curiously, the Hubble parameter
and the energy density remain finite.
Cosmological consequences of models which encounter a future quiescent singularity
(or a `Big Break' \cite{kamen03}) have been briefly discussed in \cite{as02,ss02b,kamen03}
but need to be examined in more detail.

Finally, as discussed in section \ref{sec:phantom}, Phantom models 
with $w < -1$ expand
towards a Big Rip, at which the density and all curvature invariants become
infinite. As in the case of the Big Crunch singularity, the Big Rip
will occur only in the very distant future (if it occurs at all).
For instance, if $w = {\rm constant} \geq -1.5$, $H_0 = 70$km/sec/Mpc
and $\Omega_m = 0.3$, the time to the Big Rip exceeds 22 Gyr 
\cite{star00}.

\section{Conclusions and future directions}

From the theoretical standpoint the single most important question to be asked of
dark energy is

\centerline{Is $w = -1$ ?}

\smallskip
\n
Rephrased in terms of the Statefinder diagnostic the question is:

\smallskip

\centerline{Is $\atridot/a H^3 = 1$ ?}

\smallskip
\n
If future observations do answer this question in the affirmative\footnote{\ie
if $w = -1$ is measured to satisfyingly high accuracy} then,
in all likelihood the cosmological constant is the vacuum energy, and one
will need to review the cosmological constant problem again, in order to
fathom 
why the formally infinite quantity $\tik$ is in fact so very small. 

\begin{figure}[ht]
\centering
\includegraphics[width=9cm]{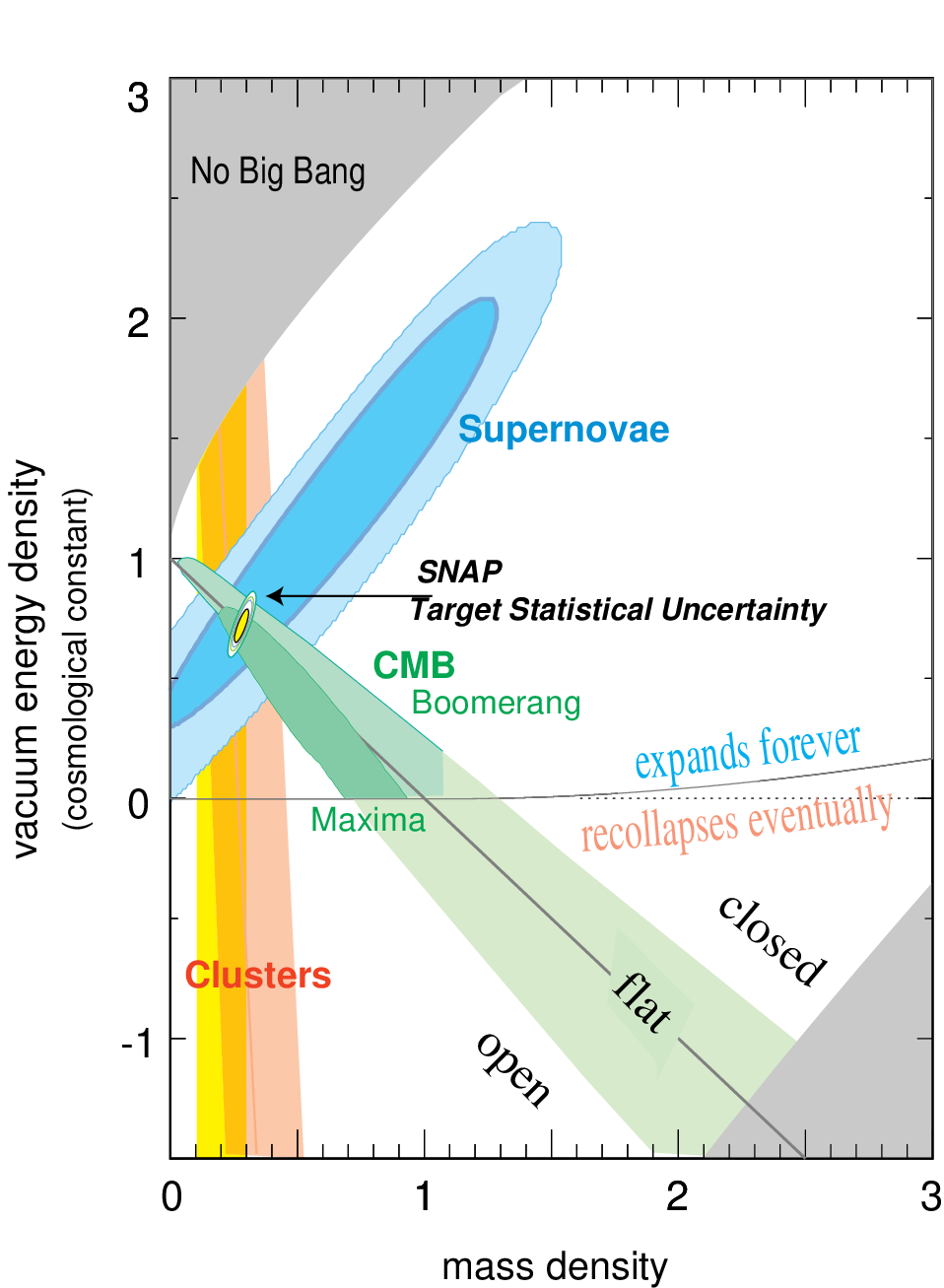}
\caption{\footnotesize
Target statistical uncertainty of the SNe experiment is shown overlaid
with current results from CMB and LSS observations. From Aldering \cite{aldering}.
}
\bigskip
\medskip
\label{fig:snap}
\end{figure}

If on the other hand, either $w \neq -1$ or if the DE density is shown to be
time dependent, then the cosmological constant problem may need to be decoupled
from the DE conundrum and searches for evolving DE models which produce
$\rho_{\rm DE} \simeq 10^{-47}$GeV$^4$ without exacerbating `cosmic coincidence'
will need to be examined deeply
in the light of developments both in high energy physics 
and in gravitation theory (superstring/M-theory, extra dimensions etc.).
In either case the key to determining the properties of
DE to great precision clearly lies with ongoing and future astrophysical 
experiments and observations.

Since the original discovery of an accelerating universe \cite{perl98a,perl98b,riess98}
the SNe data base has grown considerably and data pertaining to 
$\sim 200$ type Ia 
supernovae are avaliable in the literature
\cite{tonry03,knop03,barris03,riess04}.
Although systematic effects such as luminosity evolution or dimming
by intervening extragalactic material 
(alternatively brightening due to gravitational lensing) continue to
be a cause of some concern (recall that a luminosity evolution of
$\sim 25\%$ over a lookback time of $\sim 5$ Gyr is sufficient to nullify
the DE hypothesis \cite{riess99}),
it is reassuring that recent observations of CMB anisotropies 
and estimates of galaxy clustering in the 2dF and SDSS
surveys, make a strong and independent case for dark energy 
\cite{spergel03,tegmark03a,tegmark03b}.
Indeed, a joint analysis of CMB data from WMAP + HST Key Project determination of
$H_0$ imply $w_{\rm DE} < -0.5$ at the $95\%$ confidence level
\cite{spergel03}.

It is of paramount importance that SNe observations continue to be supplemented by other
investigations which are sensitive to the geometry of space and can be used as
independent tests of the DE hypothesis.
The volume-redshift test, Sunyaev-Zeldovich surveys, the Alcock-Paczynski
test, the angular size-redshift test and gravitational lensing have all been
suggested as possible probes of dark energy, and will doubtless enrich  
the theory vs observations debate in the near future.
In addition, the proposed 
SNe satellite which aims to measure light curves of $\sim 2000$
supernovae within a single year \cite{snap}, 
should provide a big step forward in our understanding
of type Ia supernovae and
help determine the cosmological parameters to great precision, as 
shown in figure \ref{fig:snap}.

\section{Acknowledgements}

I thank E. Papantonopoulos and the other organisers of this school
for their hospitality,
and for organising such a friendly school
in such a beautiful environment. I also thank Yuri Shtanov and Alexei Starobinsky
for their helpful comments on the first draft of this paper,
and Douglas Scott for his careful reading of the manuscript.

\section*{References}


\begin{thebibliography}{99}
\bibitem[*]{byline} E-mail: varun@iucaa.ernet.in,
\bibitem{as02}
Alam, U. and Sahni, V., astro-ph/0209443.
\bibitem{alam03a}
Alam, U., Sahni, V., Saini, T.D. and Starobinsky, A.A., \mn,
{\bf 344}, 1057 (2003)
[{\tt astro-ph/0303009}].
\bibitem{ass03}
Alam, U. Sahni, V. and Starobinsky, A.A., JCAP {\bf 0304}, 002 (2003) [astro-ph/0302302]
\bibitem{alam03b}
Alam, U., Sahni, V., Saini, T.D. and Starobinsky, A.A., astro-ph/0311364.
\bibitem{amendola99}
Amendola, L. (1999) \pd {\bf 60}, 043501.
\bibitem{amendola}
Amendola, L. and Tocchini-Valentini, D. \pd {\bf 64}, 043509 (2001);
Tocchini-Valentini, D. and Amendola, L. \pd {\bf 65}, 063508 (2002)
\bibitem{albrecht02}
Albrecht, A., Burgess, C.P., Ravndal, F. and Skordis, C., \pd {\bf 65}, 123507 (2002)
\bibitem{al_skordis}
Albrecht, A. and Skordis, C. \prl {\bf 84}, 2076 (2000).
\bibitem{chap5}
{Alcaniz}, J.S., {Jain}, D. and {Dev}, A., \pd {\bf 67}, 043514 (2003)
[{\tt astro-ph/0210476}].
\bibitem{alcaniz03}
{Alcaniz}, J.S., astro-ph/0312424
\bibitem{aldering}
Aldering, G., astro-ph/0209550
\bibitem{arbey03}
Arbey, A., Lesgourgues, J and Salati, P., \pd {\bf 68} 023511 (2003)
\bibitem{mukhanov}
{Armendariz-Picon}, C., {Mukhanov}, V. and {Steinhardt}, P.J., 
\prl {\bf 85}, 4438 (2000).
\bibitem{chap6}
{Avelino}, C., {Beca}, L.M.G., {de Carvalho}, J.P.M., Martins, C.J.A.P.
and {Pinto}, P., \pd {\bf 67}, 023511 (2003) [{\tt astro-ph/0208528}].
\bibitem{axenides04}
Axenides, M. and Dimopoulos, K., hep-th/0401238.
\bibitem{bbks86}
Bardeen, J.M., Bond, J.R., Kaiser, N. and Szalay, A.S., \apj {\bf 304}, 15 (1986)
\bibitem{dama}
Barnebei, R. et al, \plb {\bf 480}, 23 (2000)
\bibitem{exp2}
Barreiro,T., Copeland, E.J. and Nunes, N.J., \pd {\bf 61}, 127301 (2000);
\bibitem{barris03}
Barris, B. J. \etal, {\tt astro-ph/0310843}.
\bibitem{barrow86}
Barrow, J.D. and Tipler, F.J. (1986) {\em The Anthropic Cosmological
Principle}, New York: Oxford University Press.
\bibitem{barrow00}
Barrow, J.D., Bean, R. and Magueijo, J., MNRAS {\bf 316}, L41 (2000).
\bibitem{chap7} 
{Bean}, R. and {Dore}, O., \pd {\bf 68} 023515 (2003) [{\tt astro-ph/0301308}].
\bibitem{beken04}
Bekenstein, J.D., astro-ph/0403694.
\bibitem{benson02}
Benson, A.J. et al. MNRAS {\bf 333}, 177 (2002)
\bibitem{chap11}
Bento, M., Bertolami and Sen, A.A., \plb {\bf 575} 172, (2003) [astro-ph/0303538]
\bibitem{chap12}
Makler, M., Oliviera, S. and Waga, I., \pd {\bf 69}, 101301 (2004) [ astro-ph/0403378]
\bibitem{bertolami99}
Bertolami, O. and Martins, P.J. (1999) gr-qc/9910056.
\bibitem{chap2}
Bilic, N., Tupper, G.B. and Viollier, R., \plb {\bf 535}, 17 (2002).
\bibitem{bolatto02}
Bolatto, A,D, \etal \apj {\bf 565}, 238 (2002)
\bibitem{beps00}
{Boisseau}, B., {Esposito-Farese}, G., {Polarski}, D. and 
{Starobinsky}, A.A., \prl {\bf 85}, 2236 (2000)
\bibitem{bosma03}
Bosma, A., astro-ph/0312154
\bibitem{exp4}
Brax, P. and Martin, J. \pd {\bf 61}, 103502 (2000);
\plb {\bf 468}, 40 (1999).
\bibitem{brustein}
Brustein, R., in hep-ph/9810526
\bibitem{burgess03}
Burgess, C.P., Grenier, P. and Hoover, D., hep-ph/0308252;
\bibitem{burgess04}
Burgess, C.P., hep-th/0402200.
\bibitem{bullock00}
Bullock, J.S., Kravtsov, A.V. and Weinberg, D.H., \apj {\bf 539}, 517 (2000)
\bibitem{burkert95}
Burkert, A., \apj {\bf 447}, L25 (1995)
\bibitem{blok97}
de Blok, W.J.G. and McGaugh, S.S., MNRAS {\bf 290}, 533 (1997)
\bibitem{caldwell}
Caldwell, R.R., \plb {\bf 545}, 23 (2002) [{\tt astro-ph/9908168}].
\bibitem{caldwell03}
Caldwell, R.R., Kamionkowski, M. and Weinberg, N.N., 
Phys.Rev.Lett. {\bf 91} 071301 (2003) [{\tt astro-ph/0302506}].
\bibitem{capozziello03}
Capozziello, S., Carloni, S. and Troisi, A., astro-ph/0303041
\bibitem{carroll01}
Carroll, S.M., Living Rev.Rel. {\bf 4} 1 (2001) [{\tt astro-ph/0004075}].
\bibitem{carroll03}
Carroll, S.M., Hoffman, M. and Trodden, M., \pd {\bf 68}, 023509 (2003)
\bibitem{carroll03b}
Carroll, S.M., Duvvuri, V., Trodden, M. and Turner, M.S., astro-ph/0306438
\bibitem{chen00}
Chen, J.-W., Luty, M.A. and Ponton, E., JHEP {\bf 0009}, 012 (2000)
\bibitem{chiba99}
Chiba, T. \pd {\bf D60}, 083508 (1999). 
\bibitem{chiba}
Chiba, T. and Nakamura, T., \pd {\bf 62}, 121301(R) (2000).
\bibitem{chiba00}
Chiba, T., Okabe, T. and Yamaguchi, M, 2000, \pd {\bf 62}, 023511.
\bibitem{choi} Choi, K.,  
Phys.Rev. D {\bf 62} 043509 (2000) [hep-ph/9902292].
\bibitem{cline}
Cline, J.M., Jeon, S. and Moore, G.D., hep-ph/0311312.
\bibitem{coles}
Coles, P. and Lucchin, F. {\em Cosmology, The origin and evolution of cosmic
structure}, Wiley.
\bibitem{CH}
Collins,H. and and Holdom, B., Phys.\@ Rev.\@ D {\bf 62}, 105009 (2000) [{\tt
hep-ph/0003173}].
\bibitem{combes02}
Combes, F., astro-ph/0206126
\bibitem{copeland}
Copeland, E.J., Liddle, A.R. and Lidsey, J.E. 
\pd {\bf 64} 023509 (2001).
\bibitem{copeland03}
Corasaniti, P.S. and Copeland, E.J., \pd {\bf 67} 063521 (2003) [{\tt astro-ph/0205544}]
.
\bibitem{DDG}
{Deffayet}, C., {Dvali}, G. and {Gabadadze}, G., \pd {\bf 65},
044023 (2002) [{\tt astro-ph/0105068}];
{Deffayet}, C., {Landau}, S.J., {Raux}, J., {Zaldarriaga}, M. and {Astier}, P., 
\pd {\bf 66}, 024019 (2002)
[{\tt astro-ph/0201164}].
\bibitem{chap10} 
Dev, A., Jain, D. and Alcaniz, J. astro-ph/0311056
\bibitem{dimo04}
Dimopoulos, K., \pd {\bf 68}, 123506 (2003) [astro-ph/0212264]
\bibitem{dolgov03}
Dolgov, A.D. and Kawasaki, M, \plb {\bf 573}, 1 (2003) [astro-ph/0307285]
\bibitem{doran02}
Doran M. and Jaeckel, J., \pd {\bf 66}, 043519 (2003), astro-ph/0203018.
\bibitem{efstath}
Efstathiou, G., Sutherland, W. and Maddox, S.J., Nature {\bf 348}, 705 (1990).
\bibitem{elgaroy02}
Elgaroy, O. et al., \prl {\bf 90}, 021802 [hep-ph/0204152]
\bibitem{elizalde03}
Elizalde, E., Lidsey, J., Nojiri, S. and Odintsev, S.D.,
Phys. Lett. {\bf B574}, 1 (2003) [hep-th/0307177].
\bibitem{ellis03}
Ellis, J., astro-ph/0304183.
\bibitem{evans01}
Evans, N.W. astro-ph/0102082
\bibitem{chap3}
{Fabris}, J.S., {Goncalves}, S.V. and {de Souza}, P.E., {\tt astro-ph/0207430}.
\bibitem{curvaton1}
Feng, B. and Li, M., Phys.Lett., {\bf B} 564, 169 (2003) [hep-ph/0212213].
\bibitem{exp3} 
Ferreira, P.G. and Joyce, M. \prl {\bf 79}, 4740 (1997);
Ferreira, P.G. and Joyce, M. \pd {\bf 58}, 023503 (1998);
\bibitem{fischler01}
Fischler, W., Kashani-Poor, A., McNees, R. and Paban, S., JHEP 0107 003 (2001)
[hep-th/0104181].
\bibitem{flores94}
Flores, R.A. and Primack, J.R. \apj {\bf 427}, L1 (1994)
\bibitem{ft02} Frampton, P.H. and Takahashi, T.,
\plb {\bf 557}, 135 (2003) [astro-ph/0211544].
\bibitem{frampton03a}
Frampton, P., \plb {\bf 555}, 139 (2003).
\bibitem{frampton03b}
Frampton, P. and Takahashi, T., \plb {\bf 557}, 135 (2003).
\bibitem{ktf}
Freese, K. and Lewis, M., \plb {\bf 540} 1 (2002) [astro-ph/0201229]
\bibitem{frieman95}
Frieman, J., Hill, C.T., Stebbins,A. and Waga, I., \prl {\bf 75},
2077 (1995).
\bibitem{fks02}
{Frolov}, A., {Kofman}, L. and {Starobinsky}, A.A., \plb {\bf 545}, 8, (2002)
[{\tt hep-th/0204187}].
\bibitem{vilenk00}
Garriga, J. and Vilenkin, A., \pd {\bf 61} 083502 (2000).
\bibitem{efstathiou}
Gerke, B, \& Efstathiou, G., \mn {\bf 335} 33 (2002),
[{\tt astro-ph/0201336}].
\bibitem{chap4}
{Gorini}, V., {Kamenshchik}, A. and {Moschella}, U., {\tt astro-ph/0209395}.
\bibitem{kamen03}
{Gorini}, V., {Kamenshchik}, A., {Moschella}, U., {Pasquier}, V. hep-th/0311111
\bibitem{gu03}
Gu, P., Wang, X. and Zhang, X., \pd {\bf 68}, 087301 (2003)
[hep-ph/0307148].
\bibitem{hellerman01}
Hellerman, S., Kaloper, N. and Susskind, L., JHEP 0106 003 (2001) [hep-th/0104180].
\bibitem{hoekstra02}
Hoekstra, H., yee, H.K.C. and Gladders, M.D., astro-ph/0109514
\bibitem{hsu04a}
Hsu, S. and Murrey, B., astro-ph/0402541.
\bibitem{hsu04b}
Hsu, S., hep-th/0403052.
\bibitem{lidsey}
Huey, G. Huey Lidsey, J. \plb {\bf 514}, 217 (2001).
\bibitem{recon}
Huterer, D. and Turner, M. S., \pd, {\bf 60} 81301
(1999) 
\bibitem{jing00}
Jing, Y.P. and Suto, Y., apj {\bf 529}, L69 (2000)
\bibitem{johri03}
Johri, V.B., 2003, {\tt astro-ph/0311293}.
\bibitem{jungman96}
Jungman, J., Kamionkowski, M. and Griest, K., Phys. Rep. {\bf 267}, 195 (1996)
\bibitem{kl02} 
Kallosh,R., Linde, A., Prokushkin, S. and Shmakova, M.,
\pd {\bf 66} 123503 (2002) [arXiv:hep-th/0208156].
\bibitem{kl03} 
Kallosh, R. and Linde,A.
JCAP {\bf 02} 02 (2003) [astro-ph/0301087].
\bibitem{chap1}
{Kamenshchik}, A., {Moschella}, U. and {Pasquier}, V., \plb {\bf 511} 265 (2001)
[ gr-qc/0103004].
\bibitem{bridle03}
Kaplinghat, M. and Bridle, S., astro-ph/0312430
\bibitem{kauffmann93}
Kauffmann, G., White, S.D.M. and Guiderdoni, B., MNRAS {\bf 264}, 201 (1993)
\bibitem{khalil}
Khalil, S. and Munoz, C., Contemp.Phys. {\bf 43}, 51 (2002) [hep-ph/0110122].
\bibitem{khlopov}
Khlopov, M. Yu., Cosmoparticle physics, World Scientific, 1999.
\bibitem{kleina01}
Klenya, J., et al, \apj {\bf 563}, L115 (2001)
\bibitem{klypin99}
Klypin, A. et al. \apj {\bf 522}, 82 (1999)
\bibitem{knop03}
Knop, R.A., et al., 2003, {\tt astro-ph/0309368}.
\bibitem{kolb}
Kolb, E., \prl {\bf 81}, 4048 (1998)
\bibitem{kolbt90}
Kolb, E.W. and Turner, M.S. (1990) {\em The Early Universe}, Addison Wesley.
\bibitem{kolda98}
Kolda, C. and Lyth, D.H., Phys. Lett. {\bf B458} 197 (1999)
\bibitem{klypin04}
Kravtsov, A.V., Gnedin, O. and Klypin, A., astro-ph/0401088
\bibitem{li01}
Li, M., Wang, X., Feng, B. and Zhang, X.,
\bibitem{li02}
Li, M., and Zhang, X., Phys.Lett. {\bf B573}, 20 (2003) [hep-ph/0209093]
\pd {\bf 65}, 103511 (2002) [hep-ph/0112069]
\bibitem{curvaton2}
Liddle, A. R. and  Urena-Lopez, L.A., astro-ph/0302054.
\bibitem{linder}
Linder, E.V., \prl {\bf 90} 091301, (2003) [{\tt astro-ph/0208512}].
\bibitem{starkman03}
Lue, A., Scoccimaro, R. and Starkman, G.D., \pd {\bf 69}, 044005 (2004)
[{\tt astro-ph/0307034}].
\bibitem{ma04}
Ma, C.P. and Boylan-Kolchin, M., astro-ph/0403102
\bibitem{maartens03}
Maartens, R., gr-qc/0312059 
\bibitem{maartens}
Maartens, R., Wands, D., Bassett, B.A. and Heard,  I.P.C., \pd
{\bf 62}, 041301 (2000).
\bibitem{recon2}
Maor, I., Brustein, R. and Steinhardt, P.J., \prl {\bf 86}, 6 (2001)
\bibitem{maor02}
Maor, I. \etal, \pd {\bf 65} 123003 (2002) [{\tt astro-ph/0112526}].
\bibitem{maccio}
Maccio, A.V. et al., astro-ph/0309671
\bibitem{chap8}
Mackler, M., de Oliviera, Q. S. and Waga, I., \pd {\bf 68} 123521 (2003)
[astro-ph/0306507].
\bibitem{majumdar}
Majumdar, A. S. \pd {\bf 64}, 083503 (2001).
\bibitem{maroto04}
Maroto, A.L., hep-ph/0402278
\bibitem{wein98}
Martel, H., Shapiro, P., and Weinberg, S., \apj {\bf 492}, 29 (1998).
\bibitem{masaar}
Martinez, V.J. and Saar, E. ``Statistics of the galaxy distribution'',
Chapman and Hall, 2002.
\bibitem{mateo98}
Mateo, M.L., A.R.A.A. {\bf 36}, 435 (1998)
\bibitem{innes}
McInnes, B., JHEP {\bf 0208}, 029 (2002) [{\tt hep-th/0112066}].
\bibitem{milgrom}
Milgrom, M., \apj {\bf 270}, 365; {\bf 270}, 371; {\bf 270}, 384 (1983).
\bibitem{minaka02}
Minakata, H. and Sugiyama, H., hep-ph/0212240
\bibitem{mofatt04}
Mofatt, J.W., astro-ph/0403266
\bibitem{mongan01}
Mongan, T.R., Gen.Rel.Grav. {\bf 33}, 1415 (2001) [gr-qc/0103021].
\bibitem{moore99}
Moore, B. \etal, \apj {\bf 524}, L19 (1999)
\bibitem{moore99b}
Moore, B. \etal, \apj {\bf 310}, 1147 (1999)
\bibitem{chap9}
Multimaki, T., Manera. M. and Gaztanaga, E.,
\pd {\bf 69} 023004, (2004) [astro-ph/0307533]
\bibitem{munoz}
Munoz, C., IJMPA (in press) hep-ph/0309346
\bibitem{loeb03} 
Nagamine, K. and Loeb, A., New Astron. {\bf 8}, 439 (2003) [astro-ph/0204249]
\bibitem{recon1}
Nakamura, T. and Chiba, T., \mn, {\bf 306}, 696 (1999);
\bibitem{jvn02}
Narlikar, J.V. \etal, \apj {\bf 585} 1, (2003) [astro-ph/0211036], and references
therein.
\bibitem{nfw}
Navarro, J.F., Frenk, C.S. and White, S.D.M. MNRAS {\bf 275}, 720 (1995);
apj {\bf 462}, 562 (1996); apj {\bf 490}, 493 (1997)
\bibitem{leandros}
Nesseris, S. and Perivolaroupolos, L., astro-ph/0401556.
\bibitem{nojiri03a}
Nojiri, S. and Odintsev, S.D., \pd {\bf 68}, 123512 (2003)
[hep-th/0307288].
\bibitem{nojiri03b}
Nojiri, S. and Odintsev, S.D., Phys. Lett. {\bf B576}, 5 (2003)
[hep-th/0307071].
\bibitem{nojiri03c}
Nojiri, S. and Odintsev, S.D., Phys. Lett. {\bf B562}, 147 (2003)
[hep-th/0303117]
\bibitem{nojiri03d}
Nojiri, S. and Odintsev, S.D., Phys. Lett. {\bf B565}, 1, (2003)
[hep-th/0304131]
\bibitem{ngwilt} Ng, S.C. and Wiltshire, D.L., 
Phys.Rev. D {\bf 64} 123519 (2001) [astro-ph/0107142].
\bibitem{paddy03}
Padmanabhan, T., Phys. Rep. {\bf 380}, 235 (2003).
[{\tt hep-th/0212290}].
\bibitem{parker99}
Parker, L. and Raval, A., \pd {\bf 60}, 063512, 123502 (1999).
\bibitem{peebles99} 
Peebles, P.J.E. and Vilenkin, A., \pd {\bf 59} 063505 (1999).
\bibitem{pr02}
Peebles, P.J.E. and Ratra, B., Rev.Mod.Phys. {\bf 75}, 559 (2002).
[{\tt astro-ph/0207347}].
\bibitem{peri02}
Perivolaropoulos, L. and Sourdis, C., \pd {\bf 66}, 084018 (2002).
\bibitem{pbm99}
Perrotta, F., Baccigalupi, C. and Matarrese, S. \pd {\bf 61}, 023507 (1999).
\bibitem{perrotta99a}
Perrotta, F. and Baccigalupi, C. \pd {\bf 59}, 123508 (1999).
\bibitem{perrotta03}
Perrotta, F., Matarrese, S., Pietroni, M. and Schimd, C., astro-ph/0310359.
\bibitem{dark1}
Persic, M., Salucci, P. and Stel, F., astro-ph/9506004
\bibitem{phillips93}
Phillips, M.M. \apj {\bf 413}, L105 (1993).
\bibitem{pietroni03}
Peitroni, M., \pd {\bf 67}, 103523 (2003)
\bibitem{peloso03}
Peloso, M. and Poppitz, E., \pd {\bf 68}, 125009 (2003)
\bibitem{perl98a}
Perlmutter, S.J. et al., Nature {\bf 391}, 51 (1998).
\bibitem{perl98b}
Perlmutter, S.J. et al., \apj {\bf 517}, 565 (1999).
\bibitem{pope}
Pope, A.C., et al, astro-ph/0401249.
\bibitem{primack01}
Primack, J., astro-ph/0112255
\bibitem{rs}
Randall, L. and Sundrum, R., Phys. Rev. Lett. {\bf 83}, 4690 (1999).
\bibitem{ratra}
Ratra, B. and Peebels, P.J.E., \pd {\bf 37}, 3406 (1988);
\bibitem{riess99}
Riess, A.G., Filipenko, A.V., Li, W. and Schmidt, B.P. 
Astron.J. {\bf 118} 2668 (1999) [astro-ph/9907038].
\bibitem{riess98}
Riess, A.G. et al., Astron. J. {\bf 116}, 1009 (1998).
\bibitem{riess04}
Riess, A.G. et al., [astro-ph/0402512]
\bibitem{ritis99}
de Ritis, R., Marino, A.A., Rubano, C. and Scudellaro, P. 
Phys.Rev. D {\bf 62} 043506 (2000)
\bibitem{rhee03}
Rhee, G., Klypin, A. and Valenzuela, O., astro-ph/0311020
\bibitem{roszko99}
Roszkowski, L., hep-ph/9903467
\bibitem{dproy}
Roy, D.P. physics/0007025; also see E. Corbelli and
P. Salucci, astro-ph/9909252
\bibitem{sahni84}
Sahni, V., PhD thesis, Moscow State University, Moscow, 1984.
\bibitem{sc95}
Sahni, V. and Coles, P., Phys. Rept., {\bf 262}, 1 (1995).
\bibitem{sh98}
Sahni, V. and Habib, S., \prl {\bf 81}, 1766, (1998)
[{\tt hep-ph/9808204}].
\bibitem{sw00}
Sahni, V. and Wang, L., \pd {\bf 62}, 103517 (2000);
\bibitem{sami}
Sahni, V., Sami, M. and Souradeep, T., \pd {\bf 65} 023518 (2002).
\bibitem{ss02}
{Sahni}, V. and {Shtanov}, Yu.V., JCAP 0311,014, (2003) {\tt astro-ph/0202346};
\bibitem{sahni02}
Sahni, V.,
Class.Quant.Grav. {\bf 19} 3435 (2002) [astro-ph/0202076].\\
\bibitem{ss00}
Sahni, V. and Starobinsky, A.A. {\it IJMP D} {\bf 9}, 373 (2000).
\bibitem{sahni03}
Sahni, V., Saini, T.D., Starobinsky, A.A. and Alam, U., 
\jetpl {\bf 77} 201 (2003)
[{\tt astro-ph/0201498}].
\bibitem{sahni03b}
Sahni, V., Chaos, Solitons and Fractals {\bf 16}, 527 (2003)
\bibitem{saini00}
Saini, T.D., Raychaudhury, S., Sahni, V. and Starobinsky, A.A.,
\prl {\bf 85}, 1162 (2000).
\bibitem{sakharov67}
Sakharov, A.D., Dokl.\@ Akad.\@ Nauk SSSR. Ser.\@ Fiz.\@ {\bf 177}, 70 (1967) \
[Sov.\@ Phys.\@ Dokl.\@ {\bf 12}, 1040 (1968)]; \ reprinted in: Usp.\@ Fiz.\@
Nauk {\bf 161}, 64 (1991) [Sov.\@ Phys.\@ Usp.\@ {\bf 34}, 394 (1991)]; \
Gen.\@ Rel.\@ Grav.\@ {\bf 32}, 365 (2000).
\bibitem{sami2}
Sami, M., Dadhich, N.
and Shiromizu, T., Phys.Lett. B {\bf 568} 118 (2003) [hep-th/0304187]
\bibitem{sami1}
Sami, M. and Sahni, V., hep-th/0402086.
\bibitem{sand03}
Sand, D.J., Treu, T, Smith, G.P. and Ellis, R.E., astro-ph/0309465
\bibitem{sanders02}
Sanders, R.H. and McGaugh, S.S., astro-ph/0204521
\bibitem{sss96}
{Sathyaprakash}, B.S., {Sahni}, V. and
{Shandarin}, S.F., ApJL, {\bf 462}, L5 (1996)
\bibitem{bassett03}
Seery, D. and Bassett, B.A., astro-ph/0310208 
\bibitem{sss03}
Shandarin, S.F., Sheth, J.V. and Sahni, V., astro-ph/0312110
\bibitem{brane}
Shiromizu, T., Maeda, K. and Sasaki, M., \pd
{\bf 62}, 024012 (2000).
\bibitem{bi1}
Shiromizu,, T., Torii, T. and Uesugi, T., hep-th/0302223
\bibitem{singh}
Singh, P., Sami, M. and Dadhich, N.K., \pd {\bf 68}, 023522 (2003)
[{\tt hep-th/0305110}].
\bibitem{shapiro99}
Shapiro, I. and Sola, J., Phys. Lett. {\bf B475}, 236 (2000) [hep-ph/9910462]
\bibitem{shapiro03}
Shapiro, I., Sola, J., Espana-Bonet, C. and Ruiz-Lapuente, P., 
Phys. Lett. {\bf B574}, 149 (2003) [astro-ph/0303306].
\bibitem{Shtanov1}
Shtanov, Yu., hep-th/0005193.
\bibitem{ss02b}
Shtanov, Yu. and Sahni, V., Class.Quant.Grav. {\bf 19}, L101 (2003) [gr-qc/0204040]
\bibitem{dark2}
Sofue, Y. and Rubin. V., astro-ph/0010594.
\bibitem{somerville02}
Somerville, R., \apj {\bf 572} L23 (2002)
\bibitem{spergel03}
Spergel, D.N., et al, Astrophys.J.Suppl. {\bf 148}, 175 (2003) [astro-ph/0302209]
\bibitem{ss84}
Starobinsky, A. A. and Sahni, V., in {\em Modern Theoretical and Experimental Problems
of General relativity} MGPI Press, Moscow, 1984, p. 77.
\bibitem{star98}
Starobinsky, A.A., \jetpl {\bf 68}, 757 (1998).
\bibitem{star00}
Starobinsky, A.A. Grav. Cosmol. {\bf 6}, 157 (2000).
\bibitem{track1}
Steinhardt,P.J., Wang, L., and Zlatev, I., \pd {\bf 59}, 123504 (1999);
\bibitem{stein01}
Steinhardt, P.J., ``Quintessential Cosmology and Cosmic Acceleration'',
http://feynman.princeton.edu/~steinh/
\bibitem{tasi02}
Tasitsiomi, A. astro-ph/0205464
\bibitem{tegmark03a}
Tegmark, M. et al, astro-ph/0310725
\bibitem{tegmark03b}
Tegmark, M. et al, astro-ph/0310723
\bibitem{tonry03}
Tonry, J.L., et al., 2003, \apj {\bf 594}, 1, [{\tt astro-ph/0305008}].
\bibitem{um00}  Ure\~{n}a-L\'{o}pez L.A. and Matos, T. , \pd
{\bf 62}, 081302 (2000).
\bibitem{lopez}
Ure\~{n}a-L\'{o}pez, L.A. and Liddle, A., \pd {\bf 66}, 083005, (2002)
[astro-ph/0207493]
\bibitem{uzan99}
Uzan, J. \pd {\bf 59}, 123510 (1999).
\bibitem{rien}
van de Weygaert, R., 2002, ``Froth across the Universe, Dynamics and
Stochastic Geometry of
the Cosmic Foam'', arXiv: astro-ph/0206427
\bibitem{visser03}
Visser, M. gr-qc/0309109
\bibitem{wangm}
Wang. Y. and Mukherjee, P.,  astro-ph/0312192.
\bibitem{wein89}
Weinberg, S., Rev. Mod. Phys. {\bf 61}, 1 (1989).
\bibitem{wein01}
Weinberg, S., astro-ph/0104482.
\bibitem{recon3}
Weller, J. and Albrecht, A. \prl {\bf 86}, 1939 (2001).
\bibitem{albrecht}
Weller, J. and Albrecht, A., \pd {\bf 65}, 103512 (2002).
[{\tt astro-ph/0106079}].
\bibitem{wetterich88}
Wetterich, C., Nuclear Physics B {\bf 302}, 668 (1988).
\bibitem{zel68}
Zel'dovich, Ya.B., Sov. Phys. -- Uspekhi {\bf 11}, 381 (1968).
\bibitem{zhu01}
Zhu, Z.-H., Fujimoto, M.-K. and Tatsumi, D. \asta {\bf 372}, 377 (2001).
[astro-ph/0107234]
\bibitem{zimdahl03}
Zimdahl, W. and Pavon, D. gr-qc/0311067.
\bibitem{track}
Zlatev, I., Wang, L. and Steinhardt, P.J., \prl {\bf 82}, 896 (1999).
\bibitem{zwicky33}
Zwicky, F., {\em Helv. Phys. Acta} {\bf 6}, 110 (1933).
\bibitem{snap}
http://snap.lbl.gov/



\end{thebibliography}
\end{document}